\documentclass[preprint,10pt]{JHEP3} 

\JHEPspecialurl{http://jhep.sissa.it/JOURNAL/JHEP3.tar.gz}

\usepackage{epsfig,multicol,amsmath}

\def\beq{\begin{equation}}
\def\eeq{\end{equation}}
\def\bea{\begin{eqnarray}}
\def\eea{\end{eqnarray}}

\def\eq#1{{Eq.~(\ref{#1})}}
\def\fig#1{{Fig.~\ref{#1}}}
\newcommand{\bas}{\bar{\alpha}_S}
\newcommand{\as}{\alpha_S}

\newcommand{\Lb}{\left(}
\newcommand{\Rb}{\right)}

\parskip=\itemsep               

\newcommand{\h}{\frac{1}{2}}

\renewcommand{\theequation}{\thesection.\arabic{equation}}

\def\thefootnote{\fnsymbol{footnote}}
\def\blfootnote{\xdef\@thefnmark{}\@footnotetext}

\title{\huge \bf Summing  Pomeron loops in the  dipole approach}
\author{\Large  E. ~Levin  \footnotemark[4] \thanks{ leving@post.tau.ac.il,
levin@mail.desy.de;} \,\,\,\,J.~Miller \footnotemark[4]
\thanks{jeremy@post.tau.ac.il;}\,\,and\,\,\,A.~Prygarin
\footnotemark[4] \footnotemark[5]
 \hspace{0.001cm}
\thanks{ prygarin@post.tau.ac.il
} \\
\footnote{permanent address}
Department of Particle Physics, School of Physics and Astronomy\\
Raymond and Beverly Sackler
 Faculty
of Exact Science\\  Tel Aviv University, Tel Aviv, 69978, Israel
\\
\footnotemark[5] \it
Department of Particle Physics, University of Santiago de Compostela,\\
 Santiago de Compostela, 15782, Spain
}

\abstract{ In this paper we   argue that in  the kinematic range
 given by $ 1\,\,\ll\,\,\ln(1/\as^2)\,\,\ll\,\,\,\as\,Y\,\,\ll\,\,\frac{1}{\as}$,
  we can
reduce the Pomeron calculus to the exchange of non-interacting
Pomerons with the renormalized
 amplitude of their interaction with the target. Therefore, the summation of the Pomeron loops
 can be performed using the improved  Mueller, Patel,  Salam and Iancu approximation and this  leads to the
 geometrical scaling solution. This solution is found for the simplified BFKL kernel.
 We reproduce the findings of Hatta and Mueller that there are overlapping singularities. We suggest a way of dealing with these singularities.}


\preprint{  TAUP -2858-07\\
\today}

\begin{document}

\def\thefootnote{\arabic{footnote}}
\section{Introduction}
\label{sec:Int} In this paper we discuss  a solution for the  high
energy scattering amplitude in the case of high density QCD. We
would like to go beyond the mean field approximation, where the
solution has been discussed and well understood both analytically
and numerically (see  \cite{GLR,MUQI,MV,BK,JIMWLK,LT,IIM,NS}),
 and to approach
  the summation of the  so called Pomeron loops that  should be taken into account
 \cite{MSHW,LELU,IT,KOLU,HIMST}.
  The problem of taking into account the Pomeron loops
can be reduced to  the BFKL Pomeron calculus \cite{BFKL,BART,BRN}
and/or reduced to the  solution of the statistical physics problem,
i.e. the  Langevin equation and directed percolation
\cite{STPH,EGM,KLP}.

At the moment the problem of the high energy amplitude in QCD is
well understood and can be solved only in the
 kinematic region given by
\beq \label{KR}
1\,\,\ll\,\,\ln(1/\as^2)\,\,\ll\,\,\,\as\,Y\,\,\ll\,\,\frac{1}{\as}
\eeq To go beyond this region we need to know the corrections of the
order $\as^2$ to the BFKL kernel as well as the corrections
 to the vertices of the Pomeron interaction. \eq{KR} will play a significant role in our discussion and we
wish to present a more detailed derivation. It is well known that
the exchange of  one BFKL Pomeron leads to a contribution which can
be written as \beq \label{1PO} A\Lb \mbox{one Pomeron exchange}
\Rb\,\,\propto\,\,\as^2\,e^{\Delta\,Y} \eeq where $Y\,=\,\ln
(s/s_0)$ ( $s = W^2$ and $W$ is the energy of the scattering in the
c.m. frame ). The intercept of the BFKL Pomeron can be represented
in the form \beq \label{ITRC}
\Delta\,\,=\,\,\as\,\chi_{\mbox{\footnotesize{LO BFKL}}}\Lb\gamma =
1/2\Rb\,\,+\,\,\as^2\,\chi_{\mbox{\footnotesize{NLO BFKL}}}\Lb
\gamma = 1/2\Rb
 \eeq
where $\chi$ is the Mellin transform of the BFKL kernel (see more
details in \cite{KER} ).

The contribution from the exchange of two BFKL Pomerons is
proportional to \beq \label{2PO} A\Lb \mbox{two Pomeron exchange}
\Rb\,\,\propto\,\,\Lb\as^2\,e^{\Delta\,Y}\Rb^2 \eeq Comparing
\eq{1PO}  to  \eq{2PO} one can see three different kinematic
regions:
\begin{enumerate}
\item \quad $2\,\ln(1/\as)\,\,>\,\, \Delta\,Y \,\,\geq\,\,1$ ;

In this kinematic region we need to take into account the BFKL
Pomeron at leading order since $\as^2\,Y\,\,\ll\,\,1$ and neglect
the contribution from multi Pomeron exchanges.

\item \quad $1/\as\,\,\gg\,\, \Delta\,Y \,\,\gg \,\,2\,\ln(1/\as)$ ;

We need to take into account multi Pomeron exchange but we can
still  restrict ourselves to the BFKL kernel in the leading order
approximation.

\item \quad $\Delta\,Y\,\,\gg\,\,1/\as$;

In this region we have to calculate  the next-to-leading BFKL
corrections to the BFKL kernel as well as the corrections to the
Pomeron vertices.  In addition, it is not clear whether or not  we
can rely on the BFKL Pomeron calculus in this region \cite{KANCH}.
\end{enumerate}

 Concentrating on the kinematic region of \eq{KR} we develop a strategy which consists of three steps.
First, we show in this paper that the BFKL Pomeron interaction in
this kinematic region can be reduced to the exchange of
non-interacting Pomerons if we renormalize the low energy  amplitude
of the interaction of `wee' partons with the target.

Using this observation we propose an improved
Mueller-Patel-Salam-Iancu method (MPSI) for summation of Pomeron
loop diagrams. This method is actually the $t$-channel unitarity
constraints re-adjusted in a convenient form for use in the dipole
approach of QCD \cite{MUCD,MPSI}.

Finally we propose  that the answer which we obtain,  is the real
solution to our problem at ultra high energy. We will justify and
show that this is  a reasonable  hypothesis.

This paper is organised in the following way. In the next section we
discuss the simple case of the BFKL Pomeron calculus in zero
transverse dimensions. Being the simplest model for the Pomeron
interaction this approach allows us to discuss  our main  ideas and
suggestions without complex calculations. Actually, the main content
of this section has been discussed in  \cite{LEPR}, but  for
completeness we present this model in our paper. We hope that the
reader will find this instructive in later sections.

In section 3 we argue that the main properties of the BFKL Pomeron
calculus in  zero transverse dimension are inherent for the BFKL
Pomeron calculus in QCD. In particular, we can consider the
scattering amplitude in the kinematic region of \eq{KR} as the
exchange of    BFKL Pomerons neglecting their mutual interactions.

Section 4 is devoted to the calculation of the scattering amplitude
in the model for the BFKL kernel which has been developed in Ref.
\cite{KT}. In this section we demonstrate how the MPSI approach
works and we obtain a formula for the scattering amplitude. We
advocate   that the resulting scattering amplitude satisfies the
unitarity constraints both in the  $s$ and $t$ channels and could
be a good candidate for the answer
 outside   the kinematic region given by \eq{KR}.

In our conclusion we summaries our results and compare them with the
approaches that we currently have on the market.

\section{The BFKL calculus in zero transverse dimension:  non-interacting Pomerons and  improved
Mueller-Patel-Salam-Iancu approach}

\FIGURE[ht]{ \centerline{\epsfig{file=
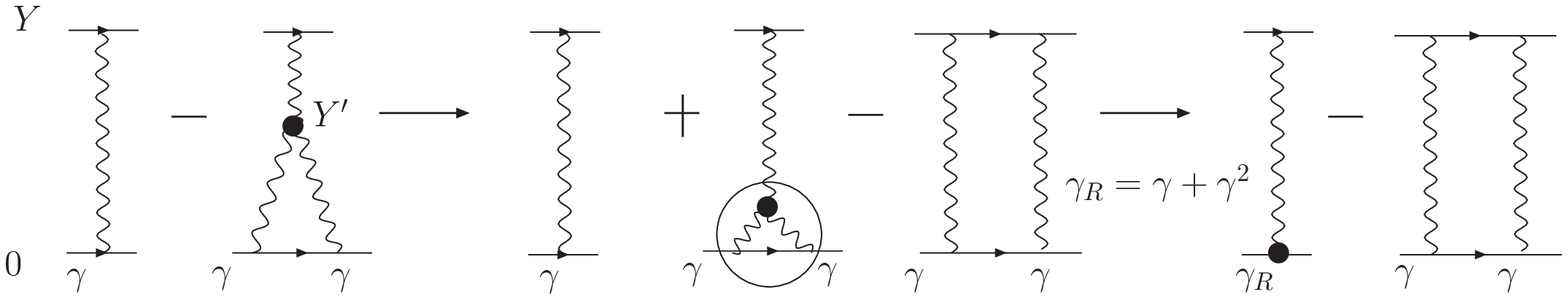,width=180mm,height=25mm}} \caption{ The renormalization
procedure  in the case of the simplest `fan' diagram. }
\label{dtrre} }

In this  section we analyse the Pomeron diagrams in the BFKL Pomeron
calculus in zero transverse dimensions (toy model of
\cite{MUCD,L1,L2}). Recently, this model has become rather popular
(see \cite{TM} and
    \cite{AMCP} where the most important aspects of this model have been discussed and solved)

Firstly we consider the simplest `fan' diagram of \fig{dtrre} and
\fig{dtrre1}. It can be calculated in an obvious way, using the
explicit expression for the Pomeron Green function,
 namely,  $G(Y - Y')\,=\,\exp\Lb \Delta\,( Y - Y')\Rb$.  Indeed, for  the diagrams of  \fig{dtrre} we have
\bea \label{MPSI1}
A\Lb\fig{dtrre}\Rb\,\,\,&=&\,\,\gamma\,G(Y-0)\,\,-\,\,\Delta\,\gamma^2\,\int^Y_0\,d\,Y'\,\,G(Y -  Y')\,G^2(Y' - 0) \,\,\,\\
 &=&\,\,\gamma\,e^{\Delta Y}\,\,\,-\,\,\,\Delta\,\gamma^2\,\int^Y_0\,d\,Y'\,\,e^{\Delta \Lb Y + Y'  \Rb}\,\\
 &=&\,\,\gamma\,e^{\Delta Y}\,\,\,-\,\,\,\Delta\,\gamma^2\,\Lb \frac{1}{\Delta}\,e^{2 \,\Delta\,Y}\,\,\,-\,\, \frac{1}{\Delta}\,e^{\Delta\,Y}\Rb\,\,\,\nonumber \\
& =&\,\,-\,\,\gamma^2\,\,e^{2
\,\Delta\,Y}\,\,\,+\,\,\,(\gamma\,+\,\gamma^2)\,\,e^{ \Delta\,Y}
\,\,=\,\,-\,\,\gamma^2\,\,e^{2
\,\Delta\,Y}\,\,\,+\,\,\gamma_{R}\,\,e^{ \Delta\,Y}
 \nonumber
 \eea
where   $\Delta$ is the Pomeron intercept which is equal to the
triple Pomeron vertex $\Gamma(1 \to 2)$ in this oversimplified
model, and $\gamma$ is the amplitude of the Pomeron interaction with
the target.

For  the diagrams of the second order, given by \fig{dtrre1},  we
have to integrate over the two rapidity variables $y_1$ and $ y_2$.
The result is
\par
$ A\Lb\fig{dtrre1}\Rb\,\,\,= $ \bea
\,\,\,\,\,&=&2\,\Delta^2\,\gamma^3\,\int^Y_0\,d\,y_1\,\,\int^{y_1}_0\,d\,y_2\,G(Y - y_1)\,G(y_1 - 0)\,G(y_2 -0)\,\,G^2(y_2 - 0) \,\, \label{MPSI2}\\
 &=&\,\,\,2\,\Delta^2\,\gamma^3\,\int^Y_0\,d\,y_1\,\int^{y_1}_0\,d\,y_2\,\,e^{\Delta \Lb Y + y_1  + y_2 \Rb}\,\,
 \,=\,\,\,2\,\Delta^2\,\gamma^3\,\Lb \frac{1}{2\,\Delta^2}\,e^{3 \,\Delta\,Y}\,\,\,-\,\, \frac{1}{\Delta^2}\,e^{2\,\Delta\,Y}\,\,+\,\, \frac{1}{2\,\Delta^2}\,e^{\,\Delta\,Y}\Rb\,\,\, \nonumber
 \eea

As one can see, the integration over $Y'$  in \fig{dtrre} and over
$y_1$ and $y_2$ in \fig{dtrre1}
 reduces these two diagrams to the sum over  diagrams which describes two contributions: the exchange
 of two non-interacting Pomerons and the exchange of one Pomeron with the renormalized vertices:
  $\gamma^{(2)}_R\,\,=\,\,\gamma\,\,+\,\,\gamma^2$ and $
\gamma^{(3)}_R\,\,=\,\,\gamma\,\,+\,\,\gamma^2\,\,+\,\,\gamma^3$

\FIGURE[ht]{
\centerline{\epsfig{file=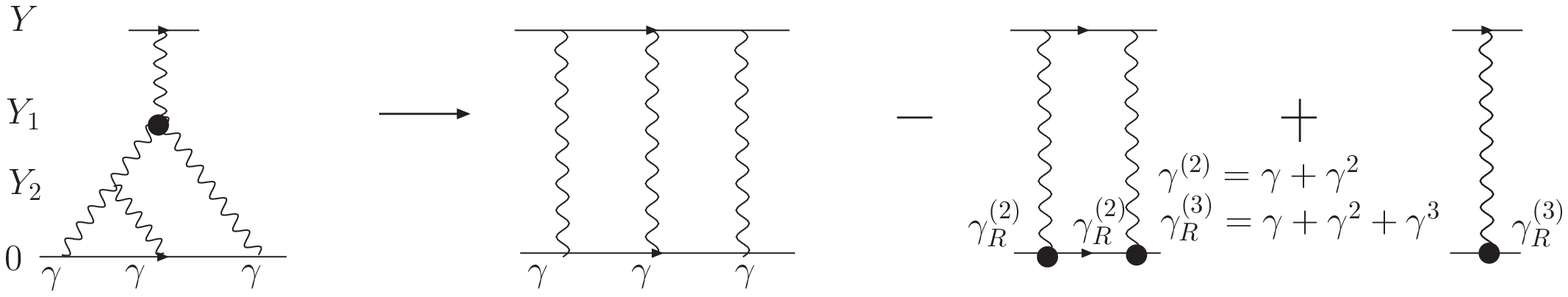,width=180mm,height=25mm}}
\caption{ The renormalization procedure  in the case of the  `fan'
diagram of the second order. } \label{dtrre1} }

Adding the two  contributions of these diagrams  gives \bea
\label{MPSI3}
&&\,A\Lb\fig{dtrre}\Rb\,\,\,+\,\,\,A\Lb\fig{dtrre1}\Rb\,\,\,\,= \\
&& =\,\,-\,\,\gamma^3\,\,e^{3
\,\Delta\,Y}\,\,\,-\,\,(\gamma^2\,+2\,\gamma^3)\,\,e^{2 \Delta\,Y}
\,\,+\,\,(\gamma \,\,+\,\,\gamma^2\,\,+\,\,\gamma^3)\,e^{3
\Delta\,Y} \, \,\,=\,\,\,\,\gamma^3\,\,e^{3
\,\Delta\,Y}\,\,\,-\,\,\,(\gamma^{(2)}_{R})^2\,\,e^{2
\Delta\,Y}\,\,+\,\,\gamma_R\,e^{\Delta\,Y} \nonumber
 \eea
Therefore, one can see that the scattering amplitude can be
rewritten as the exchange of  Pomerons without
 interaction between them but with the  renormalized   Pomeron - particle vertex. In the dipole model this
 vertex is the amplitude of the two dipole interaction in the Born approximation of perturbative QCD.

These two examples illustrate our main idea: the BFKL Pomeron
calculus in zero transverse dimensions can be viewed as the theory
of free, non-interacting Pomerons whose interaction with the target
has to be renormalized. The master equation  for the scattering
amplitude in the mean field approximation can be easily rewritten
(see Ref. \cite{LEPR}) in the form
 \beq \label{MPSI4}
  \frac{\partial
\,N_0\Lb \gamma_R|Y\Rb}{\partial\,Y}\,\,\,\,=\,\,\Gamma(1 \to
2)\,\gamma_R\,\frac{\partial \,N_0\Lb
\gamma_R|Y\Rb}{\partial\,\gamma_R}
 \eeq
 with
  \beq \label{MPSI5}
\gamma_R\,\,\,=\,\,\frac{\gamma}{1 \,\,-\,\,\gamma} \eeq

Thus we have shown in  \eq{MPSI1} and \eq{MPSI2} how  \eq{MPSI5} has
started to build up in the perturbative expansion.

The general solution of \eq{MPSI4} is the system of non -
interacting Pomerons and the scattering amplitude can be found in
the form \beq\label{MPSI6} N_0\Lb
\gamma_R|Y\Rb\,\,\,=\,\,\,\,\sum_{n=1}^{\infty}\,\,\,(-1)^n\,\,C_n\,\,\gamma^n_R\,\,G^n(Y
- 0) \eeq where the coefficients $C_n$ could be found from the
initial conditions, namely, from the expression for the low energy
amplitude. In particular, the initial condition \beq \label{MPSIIC}
 N_0\Lb \gamma_R|Y=0\Rb\,\,=\,\,\gamma\,\,=\,\,\gamma_R/(1 + \gamma_R)
\eeq
  generates $C_n\,\,=\,\,1$ and the solution is
\beq \label{MPSI7} N_0\Lb
\gamma_R|Y\Rb\,\,\,=\,\,\,\frac{\gamma_R\,e^{\Delta\,Y}}{1\,\,\,\,+\,\,\,\gamma_R\,e^{\Delta\,Y}}
\eeq The initial condition of \eq{MPSIIC} has  very simple physics
behind it that has been discussed in Ref.\cite{LEPR}.

For the analysis of the enhanced diagrams we start from the first
diagram of \fig{denre}, which can be written as follows \bea
\label{MPSI8} &&A\Lb \fig{denre}\Rb\,
    \,\,= \\
&& \,\,-\,\Delta^2\,\gamma^2\,\,\int^Y_0\,d\,y_1\,\int^{y_1}_0\,d\,y_2\,G(Y - y_1)\,G^2(y_1 - y_2)\,G(y_2 - 0) \,\,\,\nonumber \\
&&=\,\,\,-\,\Delta^2\,\gamma^2\,\,\int^Y_0\,d\,y_1\,\int^{y_1}_0\,d\,y_2\,\,e^{\Delta\,(Y
+ y_1 - y_2)}\nonumber\\
&&=\,\,-\,\,\,\gamma^2\,e^{2\,\Delta\,Y}\,\,+\,\,\gamma^2\,e^{\,\Delta\,Y}\,\,+\,\,\Delta\,\gamma^2\,Y\,
      e^{\,\Delta\,Y} \nonumber
      \eea
      where $\Gamma(2 \to 1) \,\,=\,\,\Delta\,\gamma^2$ (see Ref.\cite{IT}).

Adding \eq{MPSI8} to the exchange of  one Pomeron we obtain the
following \beq \label{MPSI9} \mbox{One Pomeron exchange} \,+\,A\Lb
\fig{denre}\Rb\,\,\,=\,\,\,\gamma_R\,e^{\,\Delta_R\,Y}
\,\,-\,\,\gamma^2\,e^{2\,\Delta\,Y} \eeq with \beq \label{MPSI10}
\gamma_R\,\,=\,\,\gamma^{(2)}\,\,=\,\,\gamma\,\,+\,\,\gamma^2\,;\,\,\,\,\,\,\,\,
\Delta_R\,\,=\,\,\Delta\,\,+\,\,\gamma\,\Delta\,; \eeq Therefore,
the Pomeron  loops  can be either large ( where their size in
rapidity is  of  the order of $Y$) and they can be considered   as
un-enhanced diagrams; or  small (where their sizes  are of the order
$1/\Delta$) and they can be treated as the renormalisation of the
Pomeron intercept.

In QCD\,\,$\Delta \,\propto \,\bas$ while
$\gamma\,\,\propto\,\as^2$. Therefore, the renormalisation of
 the Pomeron intercept $\Delta$ is proportional to $\as^3$.   We can neglect this contribution since  (i)  there a lot of $\as^2$ corrections to the kernel of the BFKL equation that  are much larger than this contribution; and (ii) in the region of $Y\,\,\ll\,\,1/\as^2$ , where we can trust our Pomeron calculus (see introduction)  $(\Delta_R\,-\,\Delta)\,Y\,\ll \,\,1$.

Concluding this analysis we can claim that the BFKL Pomeron calculus
in zero dimension is a theory of non-interacting Pomerons with
renormalised vertices of the Pomeron-particle interaction.  In the
dipole language, it means that we have a system non-interacting
Pomerons with a specific hypothesis on the amplitude
 of the dipole interactions at low energy. For the problem that we are solving here, namely, when
  we have one bare Pomeron at low energy, this amplitude is determined by \eq{MPSIIC}.

For such a system we can calculate the scattering amplitude using a
method suggested by Mueller, Patel, Salam and Iancu and developed in
a number of papers (see  Refs.\cite{MPSI,KOLE,BOR,L4,KOVG,KLTM} and
references therein). In this  method the scattering amplitude is
calculated using the
 $t$-channel unitarity constraints which is written in the following way
   (assuming that the amplitudes at high energy  are purely  imaginary, $N\,=\,Im\,A$):
\beq \label{MPSIUN} N([\dots]|Y)\,\,=\,\,N([\dots]|Y - Y';P \to
nP)\,\bigotimes \,N([\dots]|Y';P \to nP) \eeq where $\bigotimes$
stands for all necessary integrations while   $[\dots]$ describes
all quantum numbers (dipole sizes and so on ).

The correct implementation of  \eq{MPSIUN} leads for our case to the
following formula (see also Refs. \cite{KOVG,KOLE,KKLM}) \beq
\label{MPSIEQ} N^{MPSI}_0\Lb Y \Rb\,\,\,=\,\,\,1\,\,\,-\,\,\exp
\left\{\,-\,\gamma^{BA}\,\frac{\partial}{\partial
\gamma^{(1)}_{R}}\,\frac{\partial}{\partial
\gamma^{(2)}_R}\,\right\}\,N^{MFA}\Lb (\gamma^{(1)}_R|Y - Y' \Rb\,
N^{MFA}\Lb (
\gamma^{(2)}_R|Y'\Rb|_{\gamma^{(1)}_R\,=\,\gamma^{(2)}_R \,=\,0}
\eeq where $N^{MFA}\Lb  Y,\gamma_R \Rb$ is given by \eq{MPSI7}(see
also \eq{MPSI4})
  in the mean field approximation and $\gamma^{BA}\,\propto\,\as^2$ is the scattering amplitude at
   low energies which is described by the Born approximation
in perturbative QCD. The difference between \eq{MPSIEQ} and the
original MPSI approach is the fact that this equation does not
depend on the value of $Y'$  and, because of this, we do not need to
choose $Y' = Y/2$ for the best accuracy.

Substituting \eq{MPSI7} in \eq{MPSIEQ} we obtain \bea
\label{MPSIEQ1} N^{MPSI}_0\Lb \gamma^{BA}|Y\Rb\,\,\,
&=&\,\,1\,\,-\,\,\exp \Lb \frac{1}{\gamma^{BA} e^{ \Gamma(1 \to 2) Y
}}\Rb\,\frac{1}{\gamma^{BA} e^{ \Gamma( 1 \to 2)Y }}\,\, \Gamma\Lb
0,\frac{1}{\gamma^{BA} e^{ \Gamma(1 \to 2) Y } }\Rb \eea $\Gamma \Lb
0,x \Rb$ is the incomplete gamma function  (see  formulae {\bf 8.350
- 8.359} in Ref. \cite{RY}).

We claim that \eq{MPSIEQ1} is the solution to our problem. One can
easily see that $N_0\Lb \gamma|Y\Rb\,\to\,1$ at high energies in
contrast to the exact solution (see  \cite{KKLM}). The exact
solution leads to the amplitude that vanishes at high energy (see
Refs.\cite{AMCP,BMMSX}). As has been mentioned the solution depends
crucially on the initial condition for the scattering amplitude at
low energies.
 For \eq{MPSIEQ1} this amplitude is equal to
\beq \label{MPSIEQ2} N^{MPSI}_0\Lb
\gamma|Y=0\Rb\,\,\,=\,\,\sum^{\infty}_{n=1}\,(-1)^{n+1}\,\,n!\,\Lb\gamma^{BA}
\Rb^n \eeq with $\gamma^{BA}\,\propto\,\as^2$.

 \eq{MPSIEQ}  can be rewritten in a more convenient form using the Cauchy formula for the derivatives, namely,
\bea \frac{\partial^n Z^{MFA}(\gamma_R|Y)}{\partial
\gamma^n_R}\,\,&=&\,\,\,n!\,\frac{1}{2\,\pi
i}\oint_C\,\,\frac{Z^{MFA}(\gamma'_R|
Y)}{\gamma'^{n+1}_R}\,d\,\gamma'_R;\label{PL2} \eea The contour $C$
in \eq{PL2} is a circle with a small radius around $\gamma_R = 0$.
However, since the  function
 $Z$ does not  grow at large $\gamma_R$ for $n \leq 1$ we can close our contour $C$ on the singularities of the
 function $Z$. We will call this new contour $C_R$.
\bea N^{MPSI}_0\Lb  Y \Rb&=&1\,\,\,-\,\,\exp
\left\{\,-\,\gamma^{BA}\,\frac{\partial}{\partial
\gamma^{(1)}_{R}}\,\frac{\partial}{\partial
\gamma^{(2)}_R}\,\right\}\,N^{MFA}\Lb (\gamma^{(1)}_R|Y - Y' \Rb\,
N^{MFA}\Lb ( \gamma^{(2)}_R|Y'
\Rb|_{\gamma^{(1)}_R\,=\,\gamma^{(2)}_R
\,=\,0} \nonumber \\
 &=&1 - \sum^{\infty}_{n=1} \frac{\Lb -\gamma^{BA}\Rb^n}{n!}\,n!\,n!\,\,\,\frac{1}{(2\,\pi\,i)^2}\oint_{C^1_R}\!\!\!\!\!d \gamma^{(1)}_R\,\,\frac{Z^{MFA}(\gamma^{(1)}_R|Y - Y')}{(\gamma^{(1)}_R)^{n + 1}}
\,\,\,\oint_{C^2_R}\!\!\!\!d\,\gamma^{(2)}_R\,\,\frac{Z^{MFA}(\gamma^{(2)}_R|Y')}{(\gamma^{(2)}_R)^{n
+ 1}}\,\,=\,\nonumber \\
&=&\frac{1}{(2\,\pi\,i)^2}\,\,\oint\,\oint\,\frac{d\,\tilde{\gamma}^{(1)}_R}{\tilde{\gamma}^{(1)}_R}\,\,\frac{d\,
\tilde{\gamma}^{(2)}_R}{\tilde{\gamma}^{(2)}_R}
 \,\left\{ 1\,\,-\,\,\exp \Lb \frac{\tilde{\gamma}^{(1)}_R\,\,\tilde{\gamma}^{(2)}_R}{\gamma^{BA} e^{ {\cal Y} }}\Rb\,\frac{\tilde{\gamma}^{(1)}_R\,\tilde{\gamma}^{(1)}_R}{\gamma^{BA}
e^{ {\cal Y} }}\,
\Gamma\Lb 0,\frac{\tilde{\gamma}^{(1)}_R\,\tilde{\gamma}^{(2)}_R}{\gamma^{BA} e^{ {\cal Y} }} \Rb \right\}\times \nonumber \\
 & \times & Z^{MFA}\Lb \tilde{\gamma}^{(1)}_R \Rb Z^{MFA}\Lb \tilde{\gamma}^{(2)}_R \Rb \label{PL3}
\eea Here we introduce new variables
$\tilde{\gamma}^{(1)}_R\,=\,\gamma^{(1)}_R\,\exp\Lb {\cal Y}  -
{\cal Y'} \Rb$ and
 $\tilde{\gamma}^{(2)}_R\,=\,\gamma^{(2)}_R\,\exp\Lb {\cal Y'} \Rb$. In these new variables
\beq \label{PL4}
 Z^{MFA}\Lb \tilde{\gamma}^{(1)}_R \Rb\,\,=\,\,\frac{ 1 }{1\,\,+\,\,\tilde{\gamma}^{(1)}_R} \,;\,\,\,\,\,\,\,\,\,\,\,
 Z^{MFA}\Lb \tilde{\gamma}^{(2)}_R \Rb\,\,=\,\,\frac{ 1}{1\,\,+\,\,\tilde{\gamma}^{(2)}_R}
 \eeq
 Closing the integration on the poles  $ \tilde{\gamma}^{(1)}_R \,\,=\,\,- 1$ and $  \tilde{\gamma}^{(2)}_R \,\,=\,\,- 1$
we obtain the formula of \eq{MPSIEQ1}.

\FIGURE[ht]{
\centerline{\epsfig{file=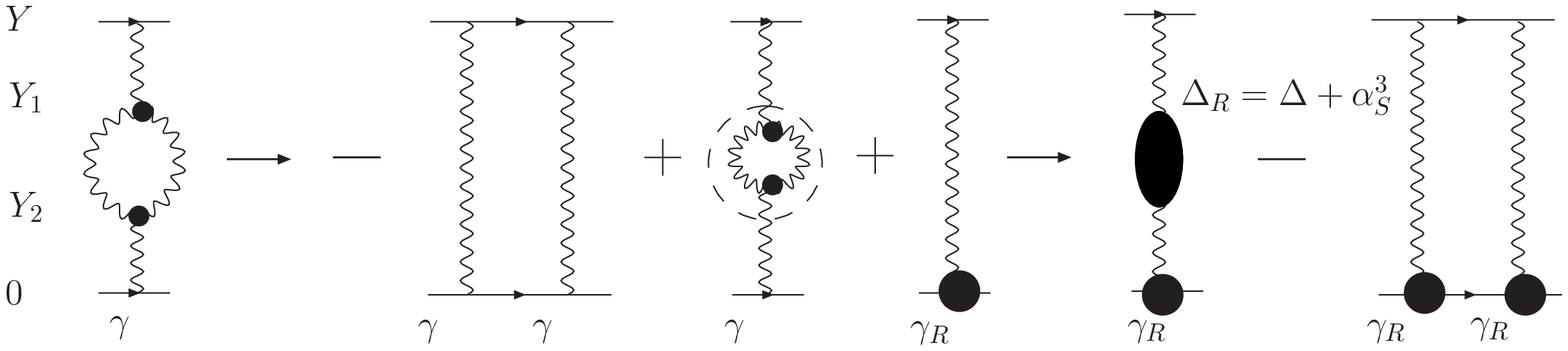,width=180mm,height=32mm}}
\caption{ The renormalization procedure  in the case of the simplest
enhanced diagram. } \label{denre} }

\section{Calculation of the simplest diagrams in the BFKL Pomeron calculus}
The main goal of this paper is to show that  in the general case of
the BFKL Pomeron calculus we have the same situation as in the
simple model.  We start from the analysis of the simple diagrams of
the BFKL Pomeron interaction.
\subsection{The functional integral formulation of the BFKL Pomeron calculus}

The most economic form of the BFKL Pomeron calculus was suggested
and developed by Braun in Ref.\cite{BRN} where he formulated this
theory as the functional integral \beq \label{FI} Z[\Phi,
\Phi^+]\,\,=\,\,\int \,\,D \Phi\,D\Phi^+\,e^S \,\,\,\mbox{with}\,S
\,=\,S_0 \,+\,S_I\,+\,S_E \eeq where $S_0$ describes free Pomerons,
$S_I$ corresponds to their mutual interaction while $S_E$ relates to
the interaction with  external sources (target and projectile).

They have the form
\bea S_0\,&=&\,\int\,d Y \,d Y'\,d^2 \underline{x}_1\, d^2
\underline{x}_2\,d^2 \underline{x}'_1\, d^2 \underline{x}'_2\, \Phi^+(\underline{x}_1,\underline{x}_2;Y)\,
G^{-1}(\underline{x}_1,\underline{x}_2;Y|\underline{x}'_1,\underline{x}'_2;Y')\,
\Phi(\underline{x}'_1,\underline{x}'_2;Y')\,; \label{S0}\\
 S_I\,&=&\,\frac{2\,\pi \bas^2}{N_c}\,\int \,d Y\,\int
\,\frac{d^2 \underline{x}_1\,d^2 \underline{x}_2\,d^2 \underline{x}_3}{\underline{x}^2_{12}
\,\underline{x}^2_{23}\,\underline{x}^2_{31}}\,
\cdot \{ \left( L_{1,2}\Phi(\underline{x}_1,\underline{x}_2;Y)\,\right) \,
\Phi^+(\underline{x}_2,\underline{x}_3;Y)\,\Phi^+(\underline{x}_3,\underline{x}_1;Y)\,\,+\,\,h.c. \}\,; \label{SI}\\
S_E\,&=&\,-\,\int \,dY\,d^2 \underline{x}_1\,d^2 \underline{x}_2\, \{
\Phi(\underline{x}_1,\underline{x}_2;Y)\,\tau_{pr}(\underline{x}_1,\underline{x}_2;Y)\,\,
+\,\,\Phi^+(\underline{x}_1,\underline{x}_2;Y)\,\tau_{tar}(\underline{x}_1,\underline{x}_2;Y)
\} \label{SE} \eea It should be stressed that \eq{SE} describes the
local interaction both in rapidity and in coordinates and
$\tau_{pr}$ ($ \tau_{tar}$) stands for the projectile and target,
respectively.

The Pomeron Green function has the form \cite{LI,NP}
\bea \label{G}
&&G\Lb \underline{x}_1,\underline{x}_2;Y|\underline{x}'_1,\underline{x}'_2;Y'\Rb\,\equiv \nonumber
\\ && \,\,\theta(Y - Y')
\sum^{\infty}_{n=-\infty} \,\int\,d \nu\,\,d^2 \underline{x}_0
\;\;e^{\omega(n,\nu) (Y -
Y')}\,(\nu^2+\frac{n^2}{4})\,\lambda(n,\nu)\,
E^*(\underline{x}'_1,\underline{x}'_2;x_0|\nu)\,E(\underline{x}_1,\underline{x}_2;\underline{x}_0|\nu)
\eea
 where the
vertices E are given by

 \bea \label{E-def}
&&E(\underline{x}_1,\underline{x}_2;\underline{x}_0|n;\nu) = (-1)^n \left(\frac{x_{12}}{x_{10}\,x_{20}}
\right)^h\,\left(\frac{x^*_{12}}{x^*_{10}\,x^*_{20}}
\right)^{\tilde{h}} \eea

with $x_{ij} = x_i - x_j $, $x_i = x_{i,x} + i x_{i,y}$
\footnote{$x_{i,x}$ and $x_{i,y}$  are x- and y-components of the
two dimensional vector $\underline{x}_i$} ,$ x^*_i = x_{i,x} - i x_{i,y} $ ; $h
= (1+n)/2 + i\nu$ and $\tilde{h} = 1 - h^*=(1-n)/2 + i\nu$.

The energy levels $\omega(n,\nu)$  are the BFKL eigenvalues

\beq \label{OM} \omega(n,\nu)\,=\,\bar{\alpha}_S \left( \psi(1) -
Re{\, \psi\left(\frac{|n| + 1}{2} + i \nu\right)} \right) \eeq
 with
$\psi(z) = d \ln \Gamma(z)/d z$ and $\Gamma(z)$ is the Euler gamma
function.

The operator $L_{12}$ in \eq{FI} is defined as \beq \label{L}
L_{1,2}\,\,=\,\,\underline{x}^4_{12}\,\underline{p}^2_{1}\,\underline{p}^2_{2}\,\,
\,\mbox{with}\,\,\underline{p}^2\,=\,-\,\nabla^2 \eeq
with $E(\underline{x}_1,\underline{x}_2;\underline{x}_0|\nu)$ and  $ \frac{1}{\lambda(n,\nu)}$  being its  eigenfunctions
and eigenvalues respectively.
It is easy to check that
 \beq \label{LACT}
L_{1,2}\, E(\underline{x}_1,\underline{x}_2;\underline{x}_0|n;\nu) \,\,=
\,\, \frac{1}{\lambda(n, \nu )}\, E(\underline{x}_1,\underline{x}_2;\underline{x}_0|n;\nu)
\eeq
with
\begin{equation} \label{BFKLLA}
\lambda(n,\nu)\,=\frac{1}{[ \frac{( n + 1)^2}{4} +  \nu^2] [\frac{( n - 1)^2}{4} +
\nu^2]}
\end{equation}

For our further discussions we rewrite \eq{E-def} in a more convenient form as
 \bea \label{E}
E(\underline{x}_1,\underline{x}_2;\underline{x}_0|n;\nu)\,&\equiv &\,E(z_{12} - \frac{x_{12}}{2},z_{12} + \frac{x_{12}}{2}|n;\nu)\,\, \nonumber \\
&=&\,\,\,(-1)^n \left(\frac{x_{12}}{(z_{12} -
\frac{x_{12}}{2})\,(z_{12}\,+\,\frac{x_{12}}{2})}
\right)^h\,\left(\frac{x^*_{12}}{(z_{12}^* -
\frac{x^*_{12}}{2})\,(z_{12}^*\,+\,\frac{x^*_{12}}{2})}
\right)^{\tilde{h}}  \hspace{1cm} \eea

where the new variable $z_{12}$ is defined by
\bea\label{z-def}
z_{12}=\frac{x_1+x_2}{2}-x_0
\eea

\subsection{Triple Pomeron vertex in momentum representation}
In this section we calculate the triple BFKL  Pomeron vertex  which
is the key ingredient of all enhanced diagrams of the type
illustrated in  \fig{denre}. We find it easier to do all the calculations in the momentum representation.
We define the Fourier transform of the vertex function in \eq{E-def} as
\bea \label{E-fourier}
g(\underline{k},\underline{q},n,\nu)\equiv \int \frac{d^2 \underline{x}_{12}}{\underline{x}_{12}^2}\;
d^2 \underline{z} \;e^{i \underline{k}\cdot \underline{x}_{12}} \;
e^{i \underline{q}\cdot \underline{z}} \;
E(\underline{x}_1,\underline{x}_2;\underline{x}_0|n;\nu)
\eea
with $\underline{x}_{12}=\underline{x}_1-\underline{x}_2$ and
 $\underline{z}=(\underline{x}_1+\underline{x}_2)/2-\underline{x}_0=\underline{b}-\underline{x}_0$.
 As it follows from \eq{E-fourier}, the momentum $\underline{k}$ is defined as the momentum conjugate to the
 dipole size. It will be shown later that the physical meaning of the momentum $\underline{q}$ is
 a momentum transferred along the Pomeron.
The explicit expression for the function $g(\underline{k},\underline{q},n,\nu)$ is found in the Appendix A.

  Using the definition of  \eq{E-fourier} we want to express
the free Pomeron Green function defined in \eq{G} in terms of the  function $g(\underline{k},\underline{q},n,\nu)$.
To do this we find an inverse Fourier transform of $g(\underline{k},\underline{q},n,\nu)$ and substitute it into
\eq{G}. The vertex function $E$ in terms of $g(\underline{k},\underline{q},n,\nu)$ reads
\bea \label{E-inverse}
E(\underline{x}_1,\underline{x}_2;\underline{x}_0|n;\nu)=\underline{x}_{12}^2 \;
\int \frac{d^2  \underline{k} }{(2\pi)^2} \;\frac{d^2 \underline{q}}{(2\pi)^2}
\;e^{-i \underline{k}\cdot \underline{x}_{12}} \;
e^{-i \underline{q}\cdot \underline{z}}\;
g(\underline{k},\underline{q},n,\nu)
\eea
With the help of \eq{E-inverse} we rewrite the free Pomeron Green function in \eq{G} as
\bea \label{G-fourier}
&&G\Lb \underline{x}_1,\underline{x}_2;Y|\underline{x}'_1,\underline{x}'_2;Y'\Rb\,\equiv \nonumber
 \\ && \,\,\theta(Y - Y')
\sum^{\infty}_{n=-\infty} \,\int\,d \nu\,\,d^2 \underline{x}_0
\;\;e^{\omega(n,\nu) (Y -
Y')}\,(\nu^2+\frac{n^2}{4})\,\lambda(n,\nu) \times
  \\
 && \underline{x}_{12}^{\prime 2} \;
\int \frac{d^2  \underline{k}' }{(2\pi)^2} \;\frac{d^2 \underline{q}'}{(2\pi)^2}
\;e^{+i \underline{k}'\cdot \underline{x}'_{12}} \;
e^{+i \underline{q}'\cdot \underline{z}'_{12}}\;
g^*(\underline{k}',\underline{q}',n,\nu)  \; \underline{x}_{12}^{ 2} \;
\int \frac{d^2  \underline{k} }{(2\pi)^2} \;\frac{d^2 \underline{q}}{(2\pi)^2}
\;e^{-i \underline{k}\cdot \underline{x}_{12}} \;
e^{-i \underline{q}\cdot \underline{z}_{12}}\;
g(\underline{k},\underline{q},n,\nu)  =\nonumber
\\
&&
\,\,\theta(Y - Y')
\sum^{\infty}_{n=-\infty} \,\int\,d \nu\,\,d^2 \underline{x}_0
\;\;e^{\omega(n,\nu) (Y -
Y')}\,(\nu^2+\frac{n^2}{4})\,\lambda(n,\nu) \times
 \nonumber \\
 && \underline{x}_{12}^{\prime 2} \;
\int \frac{d^2  \underline{k}' }{(2\pi)^2} \;\frac{d^2 \underline{q}'}{(2\pi)^2}
\;e^{+i \underline{k}'\cdot \underline{x}'_{12}} \;
e^{+i \underline{q}'\cdot \underline{b}'_{12}}\;
e^{-i \underline{q}'\cdot \underline{x}_{0}}\;
g^*(\underline{k}',\underline{q}',n,\nu)  \;\; \underline{x}_{12}^{ 2} \;
\int \frac{d^2  \underline{k} }{(2\pi)^2} \;\frac{d^2 \underline{q}}{(2\pi)^2}
\;e^{-i \underline{k}\cdot \underline{x}_{12}} \;
e^{-i \underline{q}\cdot \underline{b}_{12} }\;e^{+i \underline{q}\cdot \underline{x}_0 }\;
g(\underline{k},\underline{q},n,\nu)\nonumber
\eea
The integration over $\underline{x}_0$ brings in a delta function $(2\pi)^2\delta^{(2)} (\underline{q}-\underline{q}')$.
After the integration over the momentum $\underline{q}'$ we end up with the expression for
free Pomeron Green function in terms of $g(\underline{k},\underline{q},n,\nu)$ as follows.
\bea \label{G-fourier-full}
&&G\Lb \underline{x}_1,\underline{x}_2;Y|\underline{x}'_1,\underline{x}'_2;Y'\Rb\,\equiv \nonumber\\
&&
\,\,\theta(Y - Y')
\sum^{\infty}_{n=-\infty} \,\int\,d \nu\,\,e^{\omega(n,\nu) (Y -
Y')}\,(\nu^2+\frac{n^2}{4})\,\lambda(n,\nu) \; (2\pi)^2 \delta^{2}(\underline{q}'-\underline{q}) \times
 \\
 && \underline{x}_{12}^{\prime 2} \;
\int \frac{d^2  \underline{k}' }{(2\pi)^2} \;\frac{d^2 \underline{q}'}{(2\pi)^2}
\;e^{+i \underline{k}'\cdot \underline{x}'_{12}} \;
e^{+i \underline{q}'\cdot \underline{b}'_{12}}\;
g^*(\underline{k}',\underline{q}',n,\nu) \; \underline{x}_{12}^{ 2} \;
\int \frac{d^2  \underline{k} }{(2\pi)^2} \;\frac{d^2 \underline{q}}{(2\pi)^2}
\;e^{-i \underline{k}\cdot \underline{x}_{12}} \;
e^{-i \underline{q}\cdot \underline{b}_{12} }\;
g(\underline{k},\underline{q},n,\nu)  =\nonumber
\\
&&
\,\,\theta(Y - Y')
\sum^{\infty}_{n=-\infty} \,\int\,d \nu\,\,e^{\omega(n,\nu) (Y -
Y')}\,(\nu^2+\frac{n^2}{4})\,\lambda(n,\nu) \;  \times
 \nonumber \\
 && \underline{x}_{12}^{\prime 2}
  \;\underline{x}_{12}^{ 2}
  \;
\int \frac{d^2  \underline{k}' }{(2\pi)^2} \;\frac{d^2 \underline{k}}{(2\pi)^2}
\;\frac{d^2 \underline{q}}{(2\pi)^2}
\;e^{+i \underline{k}'\cdot \underline{x}'_{12}} \;\;e^{-i \underline{k}\cdot \underline{x}_{12}} \;
e^{-i \underline{q}\cdot ( \underline{b}_{12}-\underline{b}'_{12})}\;
g^*(\underline{k}',\underline{q},n,\nu)
g(\underline{k},\underline{q},n,\nu)  \nonumber
\eea
 The expression in \eq{G-fourier-full} clarifies the meaning of the momentum $\underline{q}$ defined in \eq{E-fourier}
 as the momentum conjugate to the difference in the impact parameters of the interacting dipoles or, alternatively,
 as the momentum transferred along the Pomeron.

 Using the same standard procedure we want to calculate the triple Pomeron vertex defined in the BFKL Pomeron Calculus
  by \eq{SI}. It is better to write the vertex in terms of free Pomeron Green functions as
  \bea \label{triple-green}
\frac{2\,\pi \bas^2}{N_c} \int dY \int
 \frac{d^2\underline{x}_1d^2\underline{x}_2d^2\underline{x}_3}
 {\underline{x}^{2}_{12}\underline{x}^{2}_{23}\underline{x}^{2}_{31}}
 (L_{12}G\Lb \underline{x}'_1,\underline{x}'_2;Y_1|\underline{x}_1,\underline{x}_2;Y\Rb)
  G\Lb \underline{x}_2,\underline{x}_3;Y|\underline{x}'_2,\underline{x}'_3;Y_2\Rb
  G\Lb \underline{x}_3,\underline{x}_1;Y|\underline{x}'_3,\underline{x}'_1;Y_3\Rb \hspace{1cm}
  \eea
  and then use the representation of
   $G\Lb \underline{x}'_1,\underline{x}'_2;Y'|\underline{x}_1,\underline{x}_2;Y\Rb$ in terms of the vertex function $E$
   defined in \eq{G}. Thus one can see that the calculation of the triple Pomeron vertex in terms of the functions
   $g(\underline{k},\underline{q},n,\nu)$ is reduced to the calculation of
   \bea \label{triple-mom-1}
\frac{2\,\pi \bas^2}{N_c}\int
 \frac{d^2\underline{x}_1d^2\underline{x}_2d^2\underline{x}_3}
 {\underline{x}^{2}_{12}\underline{x}^{2}_{23}\underline{x}^{2}_{31}}
 (L_{12}E(\underline{x}_1,\underline{x}_2;{\underline{x}_0}_1|n_1;\nu_1))
  E^*(\underline{x}_2,\underline{x}_3;{\underline{x}_0}_2|n_2;\nu_2)
  E^*(\underline{x}_3,\underline{x}_1;{\underline{x}_0}_3|n_3;\nu_3)\hspace{1cm}
  \eea
  then being properly convoluted with other necessary vertex functions $E$ as defined in \eq{G}
   to reproduce \eq{triple-green}. We use the fact  that the vertex function $E$ are eigenfunctions of the operator
   $L_{12}$ (see \eq{LACT}) to recast \eq{triple-mom-1} as
\bea \label{triple-mom-2}
\frac{2\,\pi \bas^2}{N_c}  \int
 \frac{d^2\underline{x}_1d^2\underline{x}_2d^2\underline{x}_3}
 {\underline{x}^{2}_{12}\underline{x}^{2}_{23}\underline{x}^{2}_{31}}
\frac{1}{\lambda(n_1,\nu_1)} E(\underline{x}_1,\underline{x}_2;{\underline{x}_0}_1|n_1;\nu_1)
  E^*(\underline{x}_2,\underline{x}_3;{\underline{x}_0}_2|n_2;\nu_2)
  E^*(\underline{x}_3,\underline{x}_1;{\underline{x}_0}_3|n_3;\nu_3)\hspace{1cm}
  \eea

Following the standard procedure we used above in finding the Pomeron Green function in terms of functions
$g(\underline{k},\underline{q},n,\nu)$ we insert the inverse Fourier transform of \eq{E-inverse} into
\eq{triple-mom-2} and then integrate over the coordinates $\underline{x}_1$, $\underline{x}_2$ and $\underline{x}_3$
as follows.
\bea \label{triple-mom-3}
\frac{2\,\pi \bas^2}{N_c}  \int
 \frac{d^2\underline{x}_1d^2\underline{x}_2d^2\underline{x}_3}
 {\underline{x}^{2}_{12}\underline{x}^{2}_{23}\underline{x}^{2}_{31}}
\frac{1}{\lambda(n_1,\nu_1)} \;
 &&
\underline{x}_{12}^2 \;
\int \frac{d^2  \underline{k}_{12} }{(2\pi)^2} \;\frac{d^2 \underline{q}_1}{(2\pi)^2}
\;e^{-i \underline{k}_{12}\cdot \underline{x}_{12}} \;
e^{-i \underline{q}_1\cdot \underline{z}_{12}}\;
g(\underline{k}_{12},\underline{q}_1,n_1,\nu_1) \nonumber \times \hspace{1cm}\\
  &&
  \underline{x}_{23}^2 \;
\int \frac{d^2  \underline{k}_{23} }{(2\pi)^2} \;\frac{d^2 \underline{q}_2}{(2\pi)^2}
\;e^{+i \underline{k}_{23}\cdot \underline{x}_{23}} \;
e^{+i \underline{q}_2\cdot \underline{z}_{23}}\;
g^*(\underline{k}_{23},\underline{q}_2,n_2,\nu_2) \times \hspace{1cm}\\
&&
  \underline{x}_{31}^2 \;
\int \frac{d^2  \underline{k}_{31} }{(2\pi)^2} \;\frac{d^2 \underline{q}_3}{(2\pi)^2}
\;e^{+i \underline{k}_{31}\cdot \underline{x}_{31}} \;
e^{+i \underline{q}_3\cdot \underline{z}_{31}}\;
g^*(\underline{k}_{31},\underline{q}_3,n_3,\nu_3) \nonumber =
\hspace{1cm}
\eea
\bea \label{triple-mom-4}
\frac{2\,\pi \bas^2}{N_c}  \int
 \frac{d^2\underline{x}_1d^2\underline{x}_2d^2\underline{x}_3}
 {\lambda(n_1,\nu_1)}
 &&
\int \frac{d^2  \underline{k}_{12} }{(2\pi)^2} \;\frac{d^2 \underline{q}_1}{(2\pi)^2}
\;e^{-i \underline{k}_{12}\cdot (\underline{x}_{1}-\underline{x}_{2})} \;
e^{-i \underline{q}_1\cdot \frac{\underline{x}_{1}+\underline{x}_{2}}{2}}\;
e^{+i \underline{q}_1\cdot {\underline{x}_{0}}_1}\;
g(\underline{k}_{12},\underline{q}_1,n_1,\nu_1) \nonumber \times \hspace{1cm}\\
  &&
\int \frac{d^2  \underline{k}_{23} }{(2\pi)^2} \;\frac{d^2 \underline{q}_2}{(2\pi)^2}
\;e^{+i \underline{k}_{23}\cdot (\underline{x}_{2}-\underline{x}_{3})} \;
e^{+i \underline{q}_2\cdot \frac{\underline{x}_{2}+\underline{x}_{3}}{2}}\;
e^{-i \underline{q}_2\cdot {\underline{x}_{0}}_2}\;
g^*(\underline{k}_{23},\underline{q}_2,n_2,\nu_2) \times \hspace{1cm}\\
&&
\int \frac{d^2  \underline{k}_{31} }{(2\pi)^2} \;\frac{d^2 \underline{q}_3}{(2\pi)^2}
\;e^{+i \underline{k}_{31}\cdot (\underline{x}_{3}-\underline{x}_{1})} \;
e^{+i \underline{q}_3\cdot \frac{\underline{x}_{3}+\underline{x}_{1}}{2}}\;
e^{-i \underline{q}_3\cdot {\underline{x}_{0}}_3}\;
g^*(\underline{k}_{31},\underline{q}_3,n_3,\nu_3) \nonumber \hspace{1cm}
  \eea
The integration over coordinates in \eq{triple-mom-4} gives
\bea \label{delta-1}
\begin{tabular}{ccc}
$\int d^2 \underline{x}_1$ & $\Longrightarrow$  &
$(2\pi)^2 \delta^{(2)} (\underline{k}_{12}+\frac{1}{2}\underline{q}_1+\underline{k}_{31}-\frac{1}{2}\underline{q}_3)$ \\
& &
\\
$\int d^2 \underline{x}_2$ & $\Longrightarrow$  &
$(2\pi)^2 \delta^{(2)} (-\underline{k}_{12}+\frac{1}{2}\underline{q}_1-\underline{k}_{23}-\frac{1}{2}\underline{q}_2)$ \\
& &
\\
$\int d^2 \underline{x}_3$ & $\Longrightarrow$  &
$(2\pi)^2 \delta^{(2)} (\underline{k}_{23}-\frac{1}{2}\underline{q}_2-\underline{k}_{31}-\frac{1}{2}\underline{q}_3)$
\end{tabular}
\eea
Now we perform the integration over $\underline{k}_{23}$, $\underline{k}_{31}$ and $\underline{q}_{3}$ which results into

\bea \label{delta-2}
\begin{tabular}{ccc}
$\int d^2 \underline{k}_{31}$ & $\Longrightarrow$  &
 $\underline{k}_{31}=\frac{1}{2}\underline{q}_3-\underline{k}_{12}-\frac{1}{2}\underline{q}_1$
 \hspace{0.5cm} and \hspace{0.5cm}
$(2\pi)^2 \delta^{(2)} (\frac{1}{2}\underline{q}_3-\underline{k}_{12}-\frac{1}{2}\underline{q}_1+\frac{1}{2}\underline{q}_2
-\underline{k}_{23})$
\\
& &
\\
$\int d^2 \underline{k}_{23}$ & $\Longrightarrow$  &
 $\underline{k}_{23}=\frac{1}{2}\underline{q}_3-\underline{k}_{12}-\frac{1}{2}\underline{q}_1+\frac{1}{2}\underline{q}_2$
 \hspace{0.5cm} and \hspace{0.5cm}
$(2\pi)^2 \delta^{(2)} (\underline{q}_1-\underline{q}_2-\underline{q}_3)$
\\
& &
\\
$\int d^2 \underline{q}_{3}$ & $\Longrightarrow$  &
 $\underline{q}_{3}=\underline{q}_1-\underline{q}_2$
 \end{tabular}
\eea
where we used the identity  $\int \; dx \; f(x) \; \delta(x-a) \;\delta(x-b) \;= \;f(a)\;\delta(a-b)$.
The result of \eq{delta-2} can be summarized as
\bea \label{delta-3}
\underline{k}_{31}&=&-\underline{k}_{12}-\frac{1}{2}\underline{q}_2 \nonumber \\
\nonumber \\
\underline{k}_{23}&=&-\underline{k}_{12}+\frac{1}{2}\underline{q}_1-\frac{1}{2}\underline{q}_2  \\
\nonumber \\
\underline{q}_{3}&=&\underline{q}_1-\underline{q}_2 \nonumber
\eea
Finally, we are in position to write the triple Pomeron vertex in terms of functions
$g(\underline{k},\underline{q},n,\nu)$. To do this we remove the performed integration over $\underline{x}_1$,
$\underline{x}_2$,  $\underline{x}_3$,  $\underline{k}_{31}$,  $\underline{k}_{23}$ and  $\underline{q}_3$ in
\eq{triple-mom-4}
as
was shown above by replacing $\underline{k}_{31}$, $\underline{k}_{23}$ and $\underline{q}_{3}$ with the
expressions given in \eq{delta-3}, as well as, multiply \eq{triple-mom-4} by $(2\pi)^6$ from $\delta$-functions.
Thus \eq{triple-mom-4} after the outlined integrations reads
\bea \label{triple-mom-full}
&&
  \frac{d^2  \underline{k}_{12} }{(2\pi)^2} \;\frac{d^2 \underline{q}_1}{(2\pi)^2}
  \;\frac{d^2 \underline{q}_2}{(2\pi)^2}\;
e^{+i \underline{q}_1\cdot {\underline{x}_{0}}_1-i \underline{q}_2\cdot {\underline{x}_{0}}_2
-i(\underline{q}_1-\underline{q}_2)\cdot {\underline{x}_{0}}_3}\; \times\,\Gamma\Lb {k}_{12},\underline{q}_1,\underline{q}_2 \Rb \\ \nonumber
&&\mbox{where}\\
&&
\Gamma\Lb {k}_{12},\underline{q}_1,\underline{q}_2 \Rb\,\,=\,\,
\frac{2\,\pi \bas^2}{N_c} \frac{1}
 {\lambda(n_1,\nu_1)}\,\,
g(\underline{k}_{12},\underline{q}_1,n_1,\nu_1)
g^*(-\underline{k}_{12}+\frac{1}{2}\underline{q}_1-\frac{1}{2}\underline{q}_2,\underline{q}_2,n_2,\nu_2)
g^*(-\underline{k}_{12}-\frac{1}{2}\underline{q}_2,\underline{q}_1-\underline{q}_2,n_3,\nu_3) \hspace{1cm}
 \nonumber
  \eea
The expression in \eq{triple-mom-full} represents the triple Pomeron vertex in terms of functions
$g(\underline{k},\underline{q},n,\nu)$. As it was shown on the example of the free Pomeron propagator, the triple Pomeron
vertex should be properly convoluted with all the necessary vertex functions $E$ to give the amplitude of a dipole
being scattered off two dipoles. The only missing thing is the explicit expression for
 $g(\underline{k},\underline{q},n,\nu)$ which is calculated in the Appendix A and should be plugged
 into \eq{G-fourier-full} and \eq{triple-mom-full} for obtaining the final expression.
In the region of $4k^2\gg q^2$ the function  $g(\underline{k},\underline{q},0,\nu)$ is independent of $q^2$ and has
 the form (see Appendix A)
\bea \label{g-app-c}
g(\underline{k},\underline{q},0,\nu)\approx C(\nu)(k^2)^{-\h+i\nu}
\eea
where the constant $C(\nu)$ is given by
\bea \label{c-def}
C(\nu)=\frac{\pi^2}{-i\nu}2^{2i\nu}\frac{\Gamma^2(1/2-i\nu)\Gamma^2(i\nu)}{\Gamma^2(1/2+i\nu)\Gamma^2(-i\nu)}
\eea

\subsection{The simplest `fan' diagram}

In this section we calculate the set of  diagrams shown in
\fig{dtrre}, namely,  single Pomeron exchange and the first `fan'
diagram for the triple Pomeron interaction.

\subsubsection{The single Pomeron exchange}
According to our definition of the function $g(\underline{k},\underline{q},n,\nu)$ given in \eq{E-fourier} the
momenta $\underline{k}$ and $\underline{q}$ denote momenta conjugate to the size and the impact parameter of the dipole
respectively (see also the remark after \eq{G-fourier-full}). Using this notation we readily write the expression for the
dipole with rapidity $Y$ and the transverse momentum $\underline{k}$ being scattered off the dipole
with rapidity $Y'$ and the transverse momentum $\underline{k}_0$ due to the exchange of one Pomeron.
\bea\label{1P2}
A(1P)&=&\int d\nu \;g^*(\underline{k},\underline{q},0,\nu)\;\nu^2\; \lambda(0,\nu)\;g(\underline{k}_0,\underline{q},0,\nu)
e^{\omega(\nu)(Y-Y')} \nonumber \\
&=&4^2  \;\pi^4 \int d\nu \; \frac{1}{k\; k_0} \left(\frac{k_0}{k}\right)^{2i\nu}e^{\omega_0(\nu)(Y-Y')}
= 32 \pi^6 P(k_0;k|Y-Y')
\eea
where $\omega_0(\nu)\approx 4\ln2 -14\zeta(2)\nu^2$ is the expansion of the BFKL eigenvalue in the vicinity of $\nu=0$
bringing the main contribution at high energies, and
$P(k_0;k|Y-Y')$ is well known expression for single Pomeron exchange given by
\bea \label{1P1-def}
 P(k_0;k|Y-Y') =\frac{1}{\pi} \int \frac{d\nu}{2\pi} \; \frac{1}{ k\; k_0} \left(\frac{k_0}{k}\right)^{2i\nu}e^{\omega_0(\nu)(Y-Y')}
\eea
The factor of $32\;\pi^6$ merely reflects our normalization of functions $g(\underline{k},\underline{q},0,\nu)$ defined in
\eq{E-fourier}. From \eq{1P1-def} one can see that the anomalous dimension of the gluon structure function ($xG(x,Q^2)\,\,\propto \,\,\Lb Q^2 \Rb^\gamma$) is equal to $\gamma\,\,=\,\,\h \,-\,i \nu$ since
$ P(k_0;k|Y-Y')\,\propto\,(1/k^2) x G(x=\exp(Y - Y'),k^2/k^2_0)$.

\subsubsection{The first 'fan' diagram}
The first fan diagram, for interaction of one dipole with two dipoles,  is shown in \fig{fan-a}. Using the representation for triple Pomeron vertex in
momentum space calculated in \eq{triple-mom-full} we can write the amplitude corresponding to \fig{fan-a} for $n=0$
as
\bea \label{1p2p-a}
-A(P\rightarrow 2P;\fig{fan-a})&=&\frac{2\pi \bar{\alpha}^2_s}{N_c} \int_0^Y dY'
\int \frac{d^2\underline{k}^{\prime}_{1}}{(2\pi)^2}\;\frac{d^2\underline{q}^{\prime}_{1}}{(2\pi)^2}\;
\frac{d^2\underline{q}^{\prime}_{2}}{(2\pi)^2}
 d\nu_1 \;d \nu_2 \; d\nu_3 \;(2\pi)^2 \delta^{(2)}(\underline{q}_1-\underline{q}^{\prime}_1)\;
 (2\pi)^2 \delta^{(2)}(\underline{q}_2-\underline{q}^{\prime}_2)\;\times \nonumber\\
&&
\;
 (2\pi)^2 \delta^{(2)}(\underline{q}_3-\underline{q}^{\prime}_1+\underline{q}^{\prime}_2)\;
  \frac{\nu^2_1\;\nu^2_2\;\nu^3_3\;\lambda(0,\nu_1)\;\lambda(0,\nu_2)\;\lambda(0,\nu_3)}{\lambda(0,\nu_1)}
 \;e^{\omega(\nu_1)(Y-Y')}\;\;e^{\omega(\nu_2)(Y'-0)}\times \\
 &&
\;\;e^{\omega(\nu_3)(Y'-0)}\; \;g^*(\underline{k}_{1},\underline{q}_{1},0,\nu_1)
\; g(\underline{k}^{\prime}_{1},\underline{q}^{\prime}_{1},0,\nu_1)
\; g^*(-\underline{k}^{\prime}_{1}+\frac{1}{2}\underline{q}^{\prime}_{1}
-\frac{1}{2}\underline{q}^{\prime}_{2},\underline{q}^{\prime}_{1},0,\nu_2)\;
g(\underline{k}_{2},\underline{q}_{2},0,\nu_2)\; \times \nonumber \\
&&
\; g^*(-\underline{k}^{\prime}_{1}
-\frac{1}{2}\underline{q}^{\prime}_{2},\underline{q}'_1-\underline{q}'_2,0,\nu_3)\;
g(\underline{k}_{3},\underline{q}_{3},0,\nu_3) \nonumber
\eea
where $\underline{k}^{\prime}_{2}=-\underline{k}^{\prime}_{1}+\frac{1}{2}\underline{q}^{\prime}_{1}
-\frac{1}{2}\underline{q}^{\prime}_{2}$,
$\underline{k}^{\prime}_{3}=-\underline{k}^{\prime}_{1}
-\frac{1}{2}\underline{q}^{\prime}_{2}$ and $\underline{q}'_{3}=\underline{q}'_1-\underline{q}'_2$
as it was shown in the derivation of \eq{triple-mom-full}.

\FIGURE[ht]{
\centerline{\epsfig{file=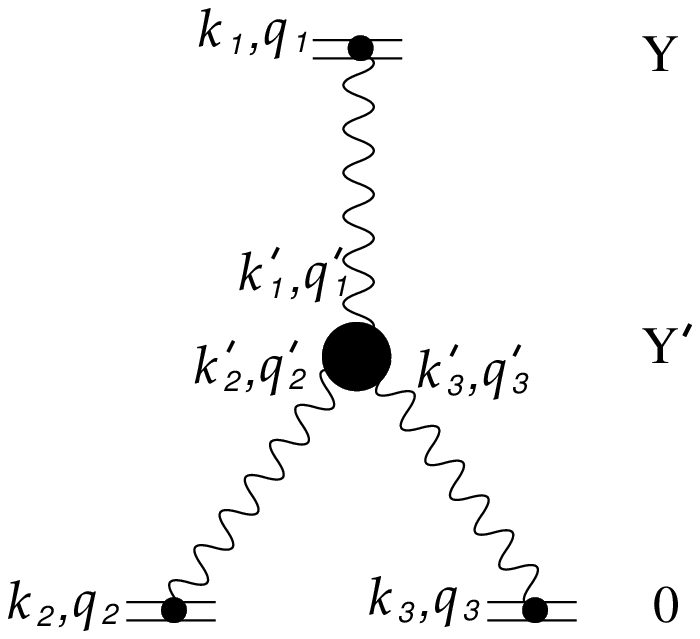,width=60mm}}
\caption{ The first 'fan' diagram. Wave lines denote the BFKL Pomerons. } \label{fan-a} }
For the case of small momenta transferred along the Pomeron we approximate the vertex function by
expression given in \eq{g-app-c}. It is worth to mention that the constants $C(\nu)$ defined in \eq{c-def}
have a property
\bea \label{c-prop}
C(\nu)C^*(\nu)=\frac{\pi^4}{\nu^2} \hspace{2cm} \hspace{2cm} \text{for}  \hspace{1cm}\nu \in \mathbb{R}
\eea
Thus \eq{1p2p-a} reads
\bea \label{1p2p-a-2}
-A(P\rightarrow 2P;\fig{fan-a})&=&\frac{2\pi \bar{\alpha}^2_s}{N_c} \int_0^Y dY'
\int \frac{d^2\underline{k}^{\prime}_{1}}{(2\pi)^2}\;\frac{d^2\underline{q}^{\prime}_{1}}{(2\pi)^2}\;
\frac{d^2\underline{q}^{\prime}_{2}}{(2\pi)^2}
 d\nu_1 \;d \nu_2 \; d\nu_3 \;(2\pi)^2 \delta^{(2)}(\underline{q}_1-\underline{q}^{\prime}_1)\;
 (2\pi)^2 \delta^{(2)}(\underline{q}_2-\underline{q}^{\prime}_2)\; \times\\
&& (2\pi)^2 \delta^{(2)}(\underline{q}_3-\underline{q}^{\prime}_1+\underline{q}^{\prime}_2)\;
(\pi^4)^3\lambda(0,\nu_2)\;\lambda(0,\nu_3)
 \; \;e^{\omega(\nu_1)(Y-Y')}\;\;e^{\omega(\nu_2)(Y'-0)}\;\;e^{\omega(\nu_3)(Y'-0)}\;\times \nonumber\\
 &&
 |\underline{k}_1|^{-1-2i\nu_1} \;{|\underline{k}'_1|}^{-1+2i\nu_1}
 \;{|-\underline{k}^{\prime}_{1}+\underline{q}^{\prime}_{1}/2
-\underline{q}^{\prime}_{2}/2|}^{-1-2i\nu_2} {|\underline{k}_2|}^{-1+2i\nu_2}\;
\;{|-\underline{k}^{\prime}_{1}
-\underline{q}^{\prime}_{2}/2|}^{-1-2i\nu_3} {|\underline{k}_3|}^{-1+2i\nu_3} \nonumber
\eea
In \eq{1p2p-a-2} momenta $\underline{q}_1$ and $\underline{q}_2$  are arbitrary. However, in the mean field approximation (MFA)  the momenta transferred along the Pomeron are much smaller
than momenta conjugate to dipole sizes  \footnote{To see this It is enough to recall that we can safely use MFA for the nucleus target where $k_1\,\approx\,k_2\,\approx\,1/R$ but $q_1 \approx q_2 \,\approx \,1/R_A$ where $R$ and $R_A$ are nucleon and nucleus radii and $R_A\,\gg\,R$.} $|\underline{k}|\gg |\underline{q}|$, which allows us to simply
\eq{1p2p-a-2} using $|-\underline{k}^{\prime}_{1}+\underline{q}^{\prime}_{1}/2
-\underline{q}^{\prime}_{2}/2|\approx |\underline{k}^{\prime}_{1}|$ and
$|-\underline{k}^{\prime}_{1}-\underline{q}^{\prime}_{2}/2|\approx |\underline{k}^{\prime}_{1}|$.
It is easy to see that the  integration over $\underline{k'}_1$ leads to
a delta function $\delta(i+2\nu_1-2\nu_2-2\nu_3)$, which is the conservation of the anomalous dimensions
($\gamma =1/2 - i\nu$) in the triple Pomeron vertex $\delta(1 + \gamma_1-\gamma_2-\gamma_3)$
(see Ref. \cite{BART} where this $\delta$-function was found for the forward scattering amplitude).
However we need to be careful since in \eq{1p2p-a-2} we implicitly used the condition that
 $k^2_3 <{k'}^2_1< k^2_1$. For  ${k'}^2_1> k^2_1$ the second term in \eq{f-full-2} gives the main contribution.
 Therefore we integrate over all the kinematic region from ${k'}^2_1=0$ to ${k'}^2_1=\infty$ but subtract
 the regions  ${k'}_1^2>k_1^2$ and  replace them by correct integral.

For example for ${k'}_1^2 > k_1^2$ we obtain the following integral

\beq \label{I1} \Lb\frac{k^2_1}{k^2_3}\Rb^{- i \nu_1}\,
\,\int^{\infty}_{l}\,d l' \,e^{ (- 1 + i2 \nu_1  -2i \nu_2 -
2i\nu_3)\,l'} \,\,-\,\Lb\frac{k^2_1}{k^2_3}\Rb^{i \nu_1}\,\,
\,\,\int^{\infty}_{l}d l' \,e^{ (- 1 -i2 \nu_1  -2i \nu_2 -
2i\nu_3)\,l'} \eeq
where $l = \ln(k_1^2/k^2_3)$.
 The result of the
integration leads to
 \beq \label{I2} e^{ (- 1  - i \nu_2-
i\nu_3)\,l}\,\,\Lb \frac{1}{- 1 + i2 \nu_1  -i 2\nu_2 -
2i\nu_3}\,\,-\,\, \frac{1}{- 1 - i2 \nu_1  -i 2\nu_2 -
2i\nu_3}\Rb \eeq
 It is easy to see that the integral over $\nu_1$ is
equal to zero. The behaviour of   $\omega(\nu_1)$ at large values of
$\nu_1$ we regularise by adding a small $i \epsilon$ term for $\nu_1$ ($
\omega(\nu_1) \to \omega(\nu_1 +i\epsilon)$. $\omega(\nu_1 +i\epsilon)\,\,
\xrightarrow{\nu_1 \to i \epsilon \pm \infty}\, - \infty$ providing a good
convergence of the integral on the large circle.

 Therefore,  finally we reproduce the
 $\delta (\gamma_1 - \gamma_2 - \gamma_3)$ contribution.

After further integration  over $\underline{q}_1$, $\underline{q}_2$ and $d\nu_1$ in \eq{1p2p-a-2} we obtain

\bea \label{1p2p-a-3}
-A(P\rightarrow 2P;\fig{fan-a})&=&\frac{2\pi \bar{\alpha}^2_s}{N_c} \int_0^Y dY' \int\; d\nu_2 \;  d\nu_3 \;
\frac{1}{2} \; (2\pi)^2 \delta^{(2)}(\underline{q}_3-\underline{q}_1+\underline{q}_2)\;
(\pi^4)^3\lambda(0,\nu_2)\;\lambda(0,\nu_3) \times \hspace{1cm} \nonumber
\\ && \;e^{\omega(i/2-\nu_2-\nu_3)(Y-Y')}\;\;e^{\omega(\nu_2)(Y'-0)}\;\;e^{\omega(\nu_3)(Y'-0)}\;
 |\underline{k}_1|^{-1-2i\nu_2-1-2i\nu_3}
 \; {|\underline{k}_2|}^{-1+2i\nu_2}\;
\; {|\underline{k}_3|}^{-1+2i\nu_3} \nonumber \\
&=&
\frac{2\pi \bar{\alpha}^2_s}{N_c}  \int\; d\nu_2 \;  d\nu_3 \;
\frac{1}{2} \; (2\pi)^2 \delta^{(2)}(\underline{q}_3-\underline{q}_1+\underline{q}_2)\;
\;
\frac{(\pi^4)^3\lambda(0,\nu_2)\;\lambda(0,\nu_3)}{\omega(-i/2+\nu_2+\nu_3)-\omega(\nu_2)-\omega(\nu_3)}
\; \times \hspace{1cm}
\\ && |\underline{k}_1|^{-1-2i\nu_2-1-2i\nu_3}
 \; {|\underline{k}_2|}^{-1+2i\nu_2}\;
\; {|\underline{k}_3|}^{-1+2i\nu_3}
\;\left(e^{\omega(-i/2+\nu_2+\nu_3)(Y-0)}
-e^{(\omega(\nu_2)+\omega(\nu_3))(Y-0)}\right)
\;\nonumber
\eea
As was already shown in the toy model in \eq{MPSI1} the integration over rapidity brings in two terms in
the last line of \eq{1p2p-a-3}. The first term corresponds to single Pomeron exchange and the second term corresponds
to double Pomeron exchange of two non-interacting Pomerons. To see this we rewrite the two terms in a more convenient
form. The first term in the last line of \eq{1p2p-a-3} reads
\bea \label{1p2p-a-4}
\frac{2\pi \bar{\alpha}^2_s}{N_c} \frac{1}{2} \;
(2\pi)^2 \delta^{(2)}(\underline{q}_3-\underline{q}_1+\underline{q}_2)\; &&\int\; d\nu_2 \;  d\mu \;
\;
\frac{(\pi^4)^3\lambda(0,\nu_2)\;\lambda(0,i/2+\mu-\nu_2)}{\omega(\mu)-\omega(\nu_2)-\omega(i/2+\mu-\nu_2)}
\; \times \hspace{1cm}
\\ && |\underline{k}_1|^{-1-2i\mu}
 \; {|\underline{k}_2|}^{-1+2i\nu_2}\;
\; {|\underline{k}_3|}^{-1+2i\mu-1-2i\nu_2}
\;e^{\omega(\mu)(Y-0)} \nonumber
\eea
where $\mu=-i/2+\nu_2+\nu_3$. At high energies the leading contribution comes from small values of $\mu$
and we can simplify \eq{1p2p-a-4} as
\bea \label{1p2p-a-5}
&&\frac{2\pi \bar{\alpha}^2_s}{N_c} \frac{1}{2} \;
(2\pi)^2 \delta^{(2)}(\underline{q}_3-\underline{q}_1+\underline{q}_2)\; \int\; d\nu_2 \;
\;
\frac{(\pi^4)^3\lambda(0,\nu_2)\;\lambda(0,-\nu_2+i/2)}{\omega(0)-\omega(\nu_2)-\omega(-\nu_2+i/2)}
\; \frac{1}{k_2k_3}\left(\frac{k_3}{k_2}\right)^{2i\nu_2}
 \int d\mu \;
\frac{1}{k_1 k_3}\left(\frac{k_3}{k_1}\right)^{2i\mu}
\;e^{\omega(\mu)(Y-0)}
=\nonumber \\
&& \hspace{3cm}
\frac{2\pi \bar{\alpha}^2_s}{N_c} \frac{1}{2} \;
(2\pi)^2 \delta^{(2)}(\underline{q}_3-\underline{q}_1+\underline{q}_2)f(k_2,k_3)
\;2\pi^2 P(k_3;k_1|Y-0)
\eea
where $P(k_3;k_1|Y-0)$ is a single Pomeron exchange defined in \eq{1P1-def} and $f(k_2,k_3)$ is
a function of $k_2$ and $k_3$ which has to be calculated and is given by
\bea \label{func-def}
f(k_2,k_3)=\int\; d\nu_2 \;
\;
\frac{(\pi^4)^3\lambda(0,\nu_2)\;\lambda(0,-\nu_2+i/2)}{\omega(0)-\omega(\nu_2)-\omega(-\nu_2+i/2)}
\; \frac{1}{k_2k_3}\left(\frac{k_3}{k_2}\right)^{2i\nu_2}
\eea

The second term of \eq{1p2p-a-3} represents two non-interacting Pomeron exchange. To see this we recast
it in the form of
\bea \label{1p2p-a-6}
&&\frac{2\pi \bar{\alpha}^2_s}{N_c}  \int\; d\nu_2 \;  d\nu_3 \;
\frac{1}{2} \; (2\pi)^2 \delta^{(2)}(\underline{q}_3-\underline{q}_1+\underline{q}_2)\;
\;
\frac{(\pi^4)^3\lambda(0,\nu_2)\;\lambda(0,\nu_3)}{\omega(-i/2+\nu_2+\nu_3)-\omega(\nu_2)-\omega(\nu_3)}
\; \times \hspace{1cm}
\\ && \frac{1}{k_1k_2}\left(\frac{k_2}{k_1}\right)^{2i\nu_2}
\;e^{\omega(\nu_2)(Y-0)}
\frac{1}{k_1k_3}\left(\frac{k_3}{k_1}\right)^{2i\nu_3}
\;e^{\omega(\nu_3)(Y-0)}
\eea
At high energies the main contribution comes from small values of $\nu_2$ and $\nu_3$ and thus we can simplify
\eq{1p2p-a-6} as follows
\bea \label{1p2p-a-7}
&&\frac{2\pi \bar{\alpha}^2_s}{N_c}
\frac{1}{2} \; (2\pi)^2 \delta^{(2)}(\underline{q}_3-\underline{q}_1+\underline{q}_2)\;
\;(4^2)^2
\frac{(\pi^4)^3}{\omega(-i/2)-\omega(0)-\omega(0)}
\; (2\pi^2)^2P(k_2,k_1|Y-0)\; P(k_3,k_1|Y-0)
\eea

 Comparing \eq{1p2p-a-3} with \eq{MPSI1} using the last  result  we conclude that
in QCD the `fan' diagrams can be viewed as the
 contribution from the exchange of two non-interacting BFKL Pomerons
  and a renormalisation of the single Pomeron contribution.

However, as it was noticed by Hatta and Mueller \cite{HM}, there is
another saddle point in \eq{1p2p-a-3}  where the denominator
 $\omega(  \nu_2 + \nu_3-i/2 )-\omega(\nu_2)\,  -\, \omega(\nu_3)$ is close to zero. Indeed,
the equation for this saddle point ( $\gamma_{SP}= 1/2 - i
\nu_{SP,1}$)  looks as follows \beq \label{FDSP}
2
\,\omega'(\gamma_{SP})\,Y\,\,-\,\,\frac{2(\omega'(2\,(\gamma_{SP}-
1)\,-\,\omega'(\gamma_{SP})}{\omega(2\,\gamma_{SP} - 1) - 2
\,\omega(\gamma_{SP})}\,  + \ln(k_1^2/k^2_3)\,\,=\,\,0 \eeq In
\eq{FDSP} we assume that the main contribution is dominated by
$\nu_1 = \nu_2$
One can see that the position of this saddle point
is very close to the solution of the equation
\beq \label{FDSP1}
\omega(2\,\gamma_0 - 1) - 2 \,\omega(\gamma_0) = 0 \eeq Denoting
$\gamma_{SP} - \gamma_0$ by $\delta \gamma$ one can see from
\eq{FDSP} that for the first term in \eq{1p2p-a-3} we have
\beq
\label{FDDG1} \delta \gamma\,=\,\frac{1}{2
\,\omega(\gamma_0)\,Y\,\,+ \,\,\ln(k_1^2/k^2_3)}\,\,\ll\,\,\gamma_0
\eeq and for the second one \beq \label{FDDG} \delta
\gamma\,=\,\frac{1}{2 \,\omega(2\gamma_0 - 1)\,Y\,\,+
\,\,\ln(k_1^2/k^2_3)}\,\,\ll\,\,\gamma_0 \eeq
The sum of these two
terms leads to the contribution with energy dependence
\bea \label{FDSP2}
&&A_{\gamma \to \gamma_0}\Lb P \to 2P; \fig{dtrre}\Rb\,\,\,\sim \,\,\,
e^{(\omega(\gamma_0 + i\delta \nu) + \omega(\gamma_0 - i \delta
\nu))\,(Y - 0)} \nonumber \eea

\DOUBLEFIGURE[ht]{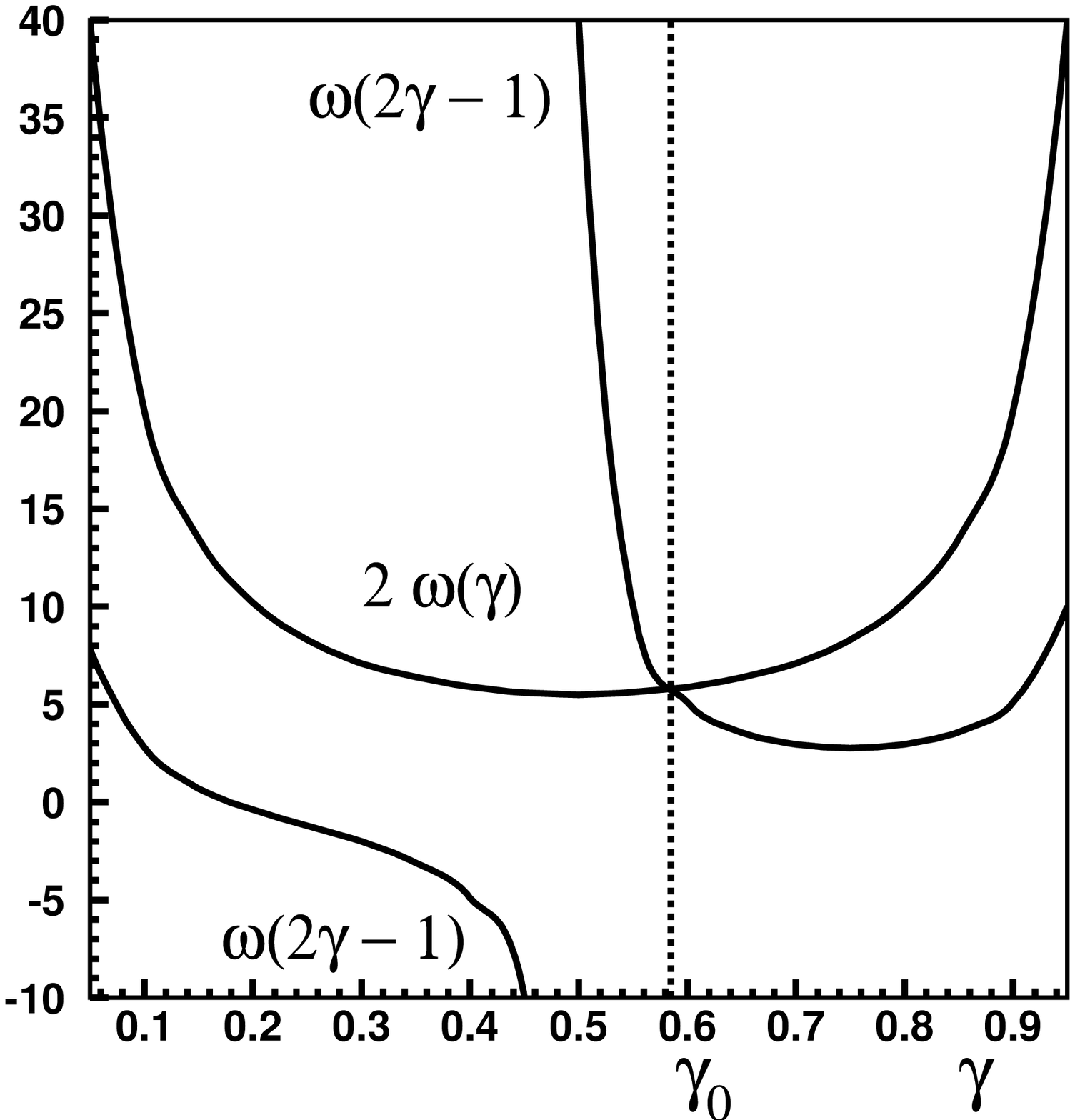,width=70mm,height=50mm}{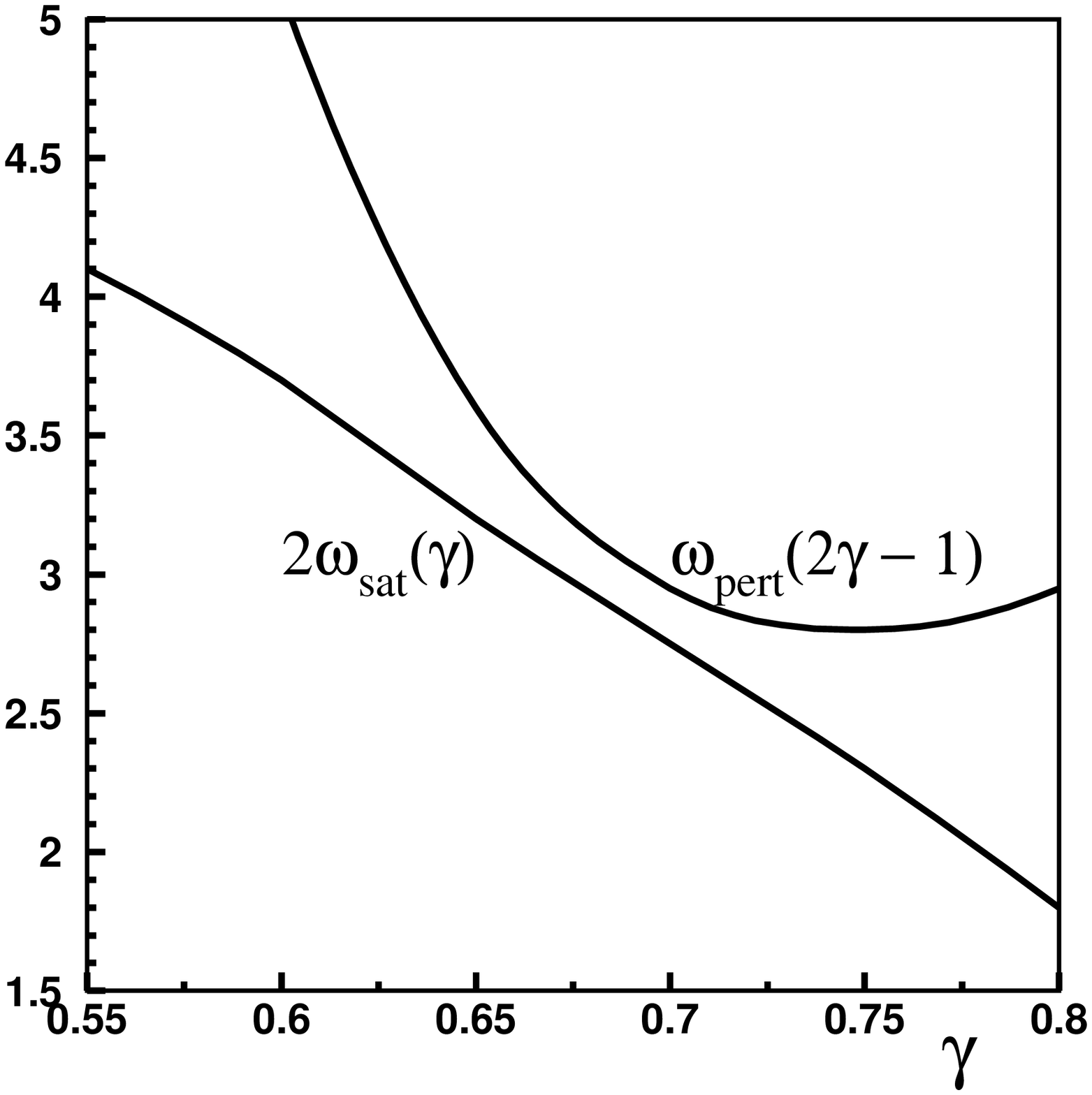,width=70mm,height=50mm}
{Solution to \protect\eq{FDSP1} for $\omega$'s, given by the BFKL
equations. \label{om0}}{ Solution to \protect\eq{FDSP1} for
$\omega_{pert}$ given by the BFKL equation and $\omega_{sat}$
calculated from \protect \eq{OMSAT}.\label{om0s}}

It turns out  (see \fig{om0}) that the resulting intercept $
2\,\omega(\gamma_0)$ is larger than the intercept for the
 two Pomeron exchange ($2 \omega(\gamma=1/2)$) and, therefore, this contribution is the largest among the three.
The appearance of such  new singularities in the angular momentum
plane  is a deadly blow  to the entire approach based on the BFKL
Pomeron. Indeed,  since the new singularity (double pole)  is
located to the right of the singularity
 generated by the exchange of two BFKL Pomerons,  we need firstly to sum over all such singularities, to obtain the resulting Green function  and only after doing this can   we  build the Reggeon calculus based on this Green function.
However, the situation changes crucially if we take into account the
fact that in \eq{FDSP1} $\omega(\gamma_0)$ and $\omega(2 \gamma_0 -
1)$  are actually in quite different kinematic regions. $\omega(2
\gamma - 1)$  enters with a small value of the argument, namely $2
\gamma_0 - 1 \,<\,\gamma_{cr}$ and, therefore, can be calculated
using the BFKL equation of  \eq{OM}. We recall that $\gamma_{cr} $
can be found from the following equation \cite{GLR,BALE,MUTR,MP}
\beq \label{GACR}
 \frac{\chi\Lb \gamma_{cr}\Rb}{ 1 -  \gamma_{cr}}\,\,=\,\,-\,\frac{d \chi\Lb \gamma_{cr}\Rb}{d
\,\gamma_{cr}} \eeq

\FIGURE[ht]{\begin{minipage}{60mm}
\centerline{\epsfig{file=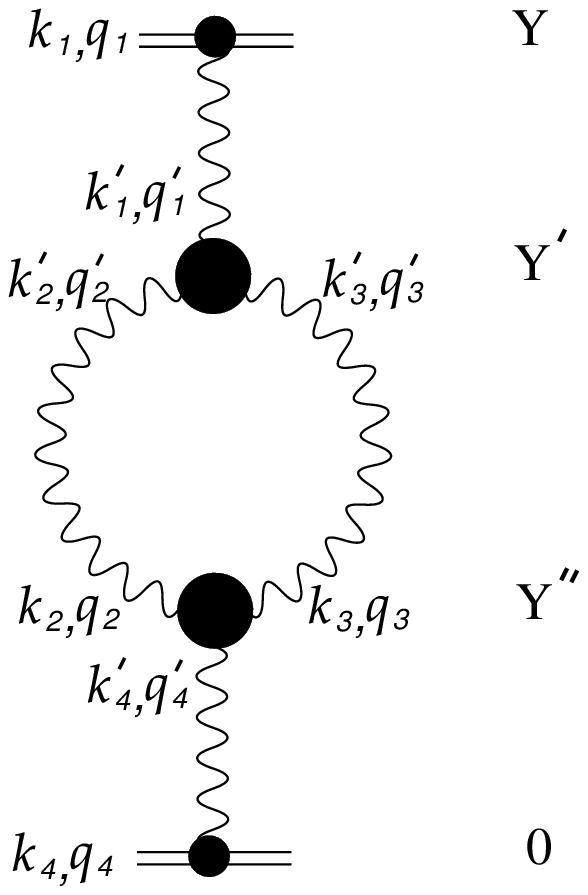,width=50mm}}
\end{minipage}
\caption{ The first enhanced diagram. } \label{1endi} }

However, we have a different situation for $\omega(\gamma_0)$.
Indeed, $\gamma_0 $ turns out to be larger than $\gamma_{cr}$ (see
\fig{om0} and recall that $\gamma_{cr} = 0.37$).  Therefore,
$\omega(\gamma_0)$ describes  the behaviour of the scattering
amplitude in the saturation region, and, therefore, cannot be
calculated using
 \eq{OM} (see also Ref.\cite{MUSH},
  where the same conclusion has been derived from  slightly different considerations).
  As it was found in Ref. \cite{BALE} for $\omega$ in the saturation region we have
\beq \label{OMSAT}
\omega_{sat}(\gamma)\,\,=\,\,\frac{\omega(\gamma_{cr})}{1 -
\gamma_{cr}}\,( 1 - \gamma) \eeq at least for  $\gamma >
\gamma_{cr}$ but close to $\gamma_{cr}$.

As one can see from \fig{om0s} \eq{FDSP1}  which  can be written as
\beq \label{FDSP1M} 2
\omega_{sat}(\gamma_0)\,\,=\,\,\omega_{pert}(2\gamma_0 - 1) \eeq has
no solution.

Having this observation in mind our suggestion is to ignore this
singularity. Our proposed  strategy is the following: we ignore all
such singularities  that appear to be due to the possibility that
two or more poles have the same position. Next we
 solve the problem and return to considering   contributions to
 the scattering amplitude having the solution both in the perturbative QCD region
 as well as in the saturation region.

Finally,  we consider only two contributions for the simplest `fan'
diagram: the first one, given by \eq{1p2p-a-3}, is due to the exchange
of two non-interacting Pomerons; and the second term of \ref{1p2p-a-3}
which is the renormalisation of the amplitude of the Pomeron
interaction with the target. Both these terms are the same as what
we considered in the BFKL Pomeron calculus in zero transverse
dimensions.

\subsection{The first enhanced diagram}

In this subsection we will calculate the simplest enhanced diagram
shown in \fig{1endi}, whose contribution has the form

\bea \label{loop-a-1}
 -A\Lb P
\to 2P  \to P; \fig{1endi}\Rb &&=
\left(\frac{2\pi\bar{\alpha}^2_s}{N_c}\right)^2 \int^Y_0 d Y^\prime
\int_0^{Y^\prime}dY^{\prime\prime} \; \int \;
\frac{d^2\underline{k}'_1}{(2\pi)^2}\frac{d^2\underline{q}'_1}{(2\pi)^2}
\frac{d^2\underline{q}'_2}{(2\pi)^2}
\frac{d^2\underline{k}'_4}{(2\pi)^2}\frac{d^2\underline{q}'_4}{(2\pi)^2}
\frac{d^2\underline{q}_3}{(2\pi)^2}
\times\\
&&\;(2\pi)^2\delta^{(2)}(\underline{q}_1-\underline{q}'_1)
\;(2\pi)^2\delta^{(2)}(\underline{q}_3-\underline{q}'_4-\underline{q}'_2)
\;(2\pi)^2\delta^{(2)}(\underline{q}_3-\underline{q}'_1-\underline{q}'_2)
\;(2\pi)^2\delta^{(2)}(\underline{q}_4-\underline{q}'_4) \times \nonumber
\\ &&
\int \;d\nu_1
 \;d\nu_2 \;d\nu_3 \;d\nu_4\; \nu^2_1 \;\nu^2_2\; \nu^2_3\;\nu^2_4\;
 \frac{\lambda(0,\nu_1)\lambda(0,\nu_2)\lambda(0,\nu_3)\lambda(0,\nu_4)}{\lambda(0,\nu_1)\lambda(0,\nu_4)}\;
 e^{\omega(\nu_1)(Y-Y')} \times \nonumber \\
 && e^{\omega(\nu_2)(Y'-Y'')}\;e^{\omega(\nu_3)(Y'-Y'')}\;e^{\omega(\nu_4)(Y''-0)}
 \; g^*(\underline{k}_1,\underline{q}_1,0,\nu_1)\;g(\underline{k}'_1,\underline{q}'_1,0,\nu_1)
 \times \nonumber \\
 && g^*(-\underline{k}'_1+\underline{q}'_1/2-\underline{q}'_2/2,\underline{q}'_2,0,\nu_2) \;
 g(-\underline{k}'_4-\underline{q}_3/2,\underline{q}_3-\underline{q}'_4,0,\nu_2)
 \times \nonumber \\
 && g^*(-\underline{k}'_1-\underline{q}'_2/2,\underline{q}'_1-\underline{q}'_2,0,\nu_3) \;
 g(-\underline{k}'_4+\underline{q}'_4/2-\underline{q}_3/2,\underline{q}_3,0,\nu_3)
 \times \nonumber \\
 && g^*(\underline{k}'_4,\underline{q}'_4,0,\nu_4) \;
 g(\underline{k}_4,\underline{q}_4,0,\nu_4) \nonumber \\
 &&=
 \left(\frac{2\pi\bar{\alpha}^2_s}{N_c}\right)^2
  (2\pi)^2\delta^{(2)}(\underline{q}_1-\underline{q}_4)
 \int^Y_0 d Y^\prime
\int_0^{Y^\prime}dY^{\prime\prime} \; \int \;
\frac{d^2\underline{k}'_1}{(2\pi)^2}
\frac{d^2\underline{q}'_2}{(2\pi)^2}
\frac{d^2\underline{k}'_4}{(2\pi)^2}
 \times
\\ &&
\int \;d\nu_1
 \;d\nu_2 \;d\nu_3 \;d\nu_4\; \nu^2_1 \;\nu^2_2\; \nu^2_3\;\nu^2_4\;
\lambda(0,\nu_2)\;\lambda(0,\nu_3)\;
 e^{\omega(\nu_1)(Y-Y')} \times \nonumber \\
 && e^{\omega(\nu_2)(Y'-Y'')}\;e^{\omega(\nu_3)(Y'-Y'')}\;e^{\omega(\nu_4)(Y''-0)}
 \;
  g^*(\underline{k}_1,\underline{q}_1,0,\nu_1)\;g(\underline{k}'_1,\underline{q}_1,0,\nu_1)
 \times \nonumber \\
 && g^*(-\underline{k}'_1+\underline{q}_1/2-\underline{q}'_2/2,\underline{q}'_2,0,\nu_2) \;
 g(-\underline{k}'_4-\underline{q}_1/2-\underline{q}'_2/2,\underline{q}_1/2+\underline{q}'_2/2-\underline{q}_4,0,\nu_2) \times
\nonumber \\
&& g^*(-\underline{k}'_1-\underline{q}'_2/2,\underline{q}_1-\underline{q}'_2,0,\nu_3) \;
 g(-\underline{k}'_4+\underline{q}_4/2-\underline{q}_1/2-\underline{q}'_2/2,\underline{q}_1/2-\underline\nu_1)
\times \nonumber \\
 && g^*(\underline{k}'_4,\underline{q}_4,0,\nu_4) \;
 g(\underline{k}_4,\underline{q}_4,0,\nu_4)\nonumber
 \eea
  where we used the fact that $\underline{k}_3=-\underline{k}'_4+\underline{q}'_4/2-\underline{q}_3/2$,
 $\underline{k}_2=-\underline{k}'_4-\underline{q}_3/2$ and
 $\underline{q}_2=\underline{q}_3-\underline{q}'_4$.

Considering, for simplicity, the case of the forward scattering ($\underline{q}_1 =  \underline{'q}_1 = \underline{q}'_4  = \underline{q}_4 = 0$ and   denoting   $\underline{q}_2 =
\underline{q}'_2   =    \underline{q}'$  as well as using the following notations:  $\nu_1 \equiv \nu$,$\nu_4 \equiv \nu'$ and $\nu_2 \equiv \nu_1$, $\nu_3 \equiv \nu_2$,
  in $\omega$ representation this diagram reduces to the following
contribution
\bea \label{FED}
-\,A\Lb \fig{1endi}\Rb &=&\Lb \frac{2 \pi \bas^2}{N_c}\Rb^2\!\!\!\int\!\nu^2 \,d\nu\,\prod^2_{i=1}\nu^2_i\,\lambda(\nu_i)\,\nu'^2\, d \nu_i \,d \nu'\,\,d^2 q'\,d^2 k_1\,d^2 k_2 \nonumber \\
 &\times &\,g^*\Lb \underline{k},0,0,\nu  \Rb\,\,g\Lb \underline{k}_1,0,0, \nu  \Rb\
  g^*(-\underline{k}'_1 + \underline{q}'/2,\underline{q}',0,\nu_1)\,
  g^*(-\underline{k}'_1 -\underline{q}'/2,-\underline{q}',0,\nu_2)\,\nonumber \\
 &\times &g(-\underline{k}_2 + \underline{q}'/2,\underline{q}',0,\nu_1)\,
 g^*(-\underline{k}_2 -\underline{q}'/2,-\underline{q}',0,\nu_2)\,
g^*\Lb \underline{k}_4,0,0,\nu'  \Rb\,\,g\Lb \underline{k}_4,0,0, \nu'  \Rb\
\nonumber\\
&\times& \frac{1}{2\pi i}\,\int^{a + i \infty}_{a - i
\infty}\!\!\!\!\!\!d\,\omega\,\,\,e^{\omega\,Y}\frac{1}{\omega -
\omega(\nu)}\,\frac{1}{\omega - \omega(\nu_1) - \omega(\nu_2)}\,
\frac{1}{\omega - \omega(\nu')}
\eea
The integral over
$\omega$ can be rewritten as the sum over different contributions.
We will show below that $\nu =\nu'$ and the integral has the form
\bea \label{OMI}
&&\frac{1}{2\pi i}\,\int^{a + i \infty}_{a - i \infty}\!\!\!\!\!\!d\,\omega\,\,\,e^{\omega\,Y}\frac{1}{\Lb \omega - \omega(\nu)\Rb^2}\,\frac{1}{\omega - \omega(\nu_1) - \omega(\nu_2)}\, \rightarrow \, \\
&&e^{\omega(\nu)\,Y}\,\Lb Y - \frac{1}{\omega(\nu) - \omega(\nu_1) -
\omega(\nu_2)} \Rb \,\frac{1}{\omega(\nu) - \omega(\nu_1) -
\omega(\nu_2)} \,\,+\,\,e^{(\omega(\nu_1) +
\omega(\nu_2))\,Y}\,\frac{1}{(\omega (\nu) - \omega(\nu_1) -
\omega(\nu_2))^2} \nonumber \eea

The first term leads to the renormalisation  of the BFKL Pomeron
intercept (the first term in the brackets) as well
 as of the vertex of the interaction of the BFKL Pomeron with the target (see \fig{denre}). The second term reduces to the exchange of two BFKL Pomerons without any interaction between them.
However, for $\gamma_0 =  \h - i \nu_0$, given by \eq{FDSP1},
instead of \eq{OMI}, we have the pole of the third order which leads
to the contribution
 \bea \label{OMI1}
&&\frac{1}{2\pi i}\,\int^{a +
i \infty}_{a - i
\infty}\!\!\!\!\!\!d\,\omega\,\,\,e^{\omega\,Y}\frac{1}{\Lb \omega -
\omega(\nu)\Rb^2}\,\,\,\frac{1}{\omega - \omega(\nu_1) -
\omega(\nu_2)}\, \,=\,\,  \h\, Y^2 \,\,e^{\omega(\nu_0)\,Y}\, \eea
We will not consider this contribution by the same reason  that we
used in calculating the `fan' diagram since in this diagram two
Pomerons with $\nu$ and $\nu'$ turn out to be outside of the
saturation region while two Pomerons with $\nu_1$ and $\nu_2$ are
 located inside the saturation region where we cannot use the BFKL kernel to
 determine the values of their
intercepts. Therefore, the first `fan' and enhanced diagrams are
very similar with respect to the integration over rapidity. However,
in the case of the enhanced diagram we have an additional problem:
the $q'$ integration. Indeed, even if we assume that $k\,>k_0$ and
$q =0$,  the integration over $q'$ is restricted by the smallest of
the two momenta $k'_1$ and $k_2$(see Appendix B for more details on $q'$ integration).
  In other words we have the following region of integration:
\begin{enumerate}
\item \quad $k > k_1 > k_2 >k_0$.

In this region the first term in \eq{f-full-2} contributes for all four
Pomerons and we have the product of two $\delta$ functions in the
vertices, namely, $\delta( 1 + \gamma - \gamma_1 -
\gamma_2)\,\delta( 1 + \gamma' - \gamma_1 - \gamma_2) $. These leads
to $\gamma = \gamma' (\nu =\nu')$  and the entire contribution looks
the same
 as in the case of the first `fan' diagram. It is worthwhile mentioning that
 for dipoles inside  one BFKL Pomeron we have the same ordering in momenta.
\item \quad $k > k_1> k_0$  but $k_1 < k_2 $ while $k_2 > k_0$.
In this kinematic region the first Pomeron with $\nu$ and the fourth
one with $\nu'$  have
 contributions which stem from the first term in \eq{f-full-2}. However,  for two Pomerons with $\nu_1$ and $\nu_2$ the second terms in \eq{f-full-2} play the most important roles. Collecting all factors of $ k_1$ one can see that the integration over $k_1$ has the form
\beq \label{IK1} \int\,d\,l\,\exp\Lb ( 1 - \h - i\nu - i\nu_1 - i
\nu_2 )\,l \Rb \eeq where $l = \ln (k^2_1/k^2_0)$. The first term
(1) in the bracket
 of \eq{IK1} stemming from the integration over $q_1 \leq k^1_1$,  the second term ($ - \h - i\nu$)
 originates from the contribution of the Pomeron with  $\nu$,
 the third as well as the fourth terms reflects the product of the factor $(k^2_1)^{-\h + i \nu_1}$
 from the upper vertex and factor $(k^2_1)^{
- 2 i \nu_1}$ from the lower vertex since $q^2_1 = k^2_1$. The
integration generates $\delta(2 - \gamma - \gamma_1 - \gamma_2)$.
 The integration over $k_2$ has the same character as for the case of the first term  and leads to
  $\delta (\gamma' - \gamma_ 1  - \gamma_2)$. These $\delta$ functions give $1 - \gamma = \gamma'$.
   The decomposition of \eq{OMI} gives
once more the terms shown in \fig{denre} and  the pole of the third
order of \eq{OMI1} since $\omega(1 - \gamma) = \omega(\gamma)$ (see
\eq{OM}). There is only one  difference: in this pole of the third
order, three Pomerons (with $\nu,\nu_1$ and $\nu_2$)  are inside of
the saturation region. This fact does not influence  our reasoning
that we need not consider this kind of contribution.
\end{enumerate}

Finally, we can conclude that the first enhanced diagram has the
same three contributions as the BFKL Pomeron calculus in zero
transverse dimensions.

\subsection{Reduction of the emhanced diagrams to the system of non-interacting Pomerons}

In this section we will prove   that an arbitrary enhanced diagram (see the  diagram in \fig{dtrre}-1) can be reduced to the system of non-interacting Pomerons in the kinematic region:  $ \as Y \,\ll\,1/\as$.  As far as $\omega$ integration is concerned this diagram can be written as
\beq \label{R1}
\int^{ a + i \infty}_{a - i \infty}\,\,\frac{d \omega}{2 \pi i}\,e^{\omega\,Y}\,\,
\gamma^2_B  \,\Lb \frac{1}{\omega\,-\,\omega(\gamma)}\Rb^{m + 1}\,\Sigma^m_1\Lb 1PI|\omega\Rb
\eeq
where  $\gamma_B$ is the vertex of  Pomeron interaction with the dipole \footnote{In this section we use for this vertex the notation $\gamma_B$ instead $\gamma$ since we reserve $\gamma$ for the anomalous dimension. We hope that in other part of the paper we use $\gamma$ for the vertex will not lead to any misunderstanding.}
 $\Sigma\Lb 1PI|\omega\Rb$ is the one Pomeron irreducable diagram and  can be written in the following form
\beq \label{R2}
\Sigma\Lb 1PI|\omega\Rb\,\,=\,\,-\frac{\Gamma^2}{\omega - \omega(\gamma_1) - \omega(\gamma_2)}\,\,+ \,\,\sum_{j}\,(-1)^j\,\Gamma^j\,\prod^{j - 1}_{i>2}\,\frac{1}{\omega - \sum^i_{l,l>2}\,\omega(\gamma_l)}
\eeq
In \eq{R1} and \eq{R2} $\gamma$ denotes the Pomeron - dipole vertex and $\Gamma$ is the triple Pomeron vertex (see \eq{triple-mom-full}). The integral over  $\omega$ in \eq{R1} can be taken  closing contour over one Pomeron poles $\omega= \omega(\gamma)$  ( see the diagram in  \fig{dtrre}-2) and  closing contour over singularities of $\Sigma\Lb 1PI|\omega\Rb $ - the  diagram in \fig{dtrre}-3. The first contribution has the form
\beq \label{R3}
A(\fig{dtrre}-2)\,\,\propto\,\,\frac{1}{m!}\,\Lb Y \,\,+\,\,O\Lb 1/\as \Rb   \Rb^m\,\exp \Lb \omega(\gamma) Y \Rb \,\Sigma^m_1\Lb 1PI|\omega=\omega(\gamma)\Rb
\eeq
The contributions of the order of $1/\as$ stem from the differentiation
of the factor $\Sigma^m_1\Lb 1PI|\omega\Rb$ with respect to $\omega$ in \eq{R1} in taking the pole of $m+1$-order.
 All these contributions are so small that they can be neglected, since $Y \,\gg\,1/\as$ ($ \as Y \gg 1$).

From \eq{R2} one can see that the first term in \, $\Sigma_1\Lb 1PI|\omega\Rb$  is of the order of $\Gamma^2/\omega(\gamma)\,\,\propto\,\as^3$ while the others lead to smaller contribution.   Therefore,  the only term with $m = 0$ remains that gives the   exchange of one Pomeron (see \fig{dtrre}-4).

\FIGURE[ht]{ \centerline{\epsfig{file=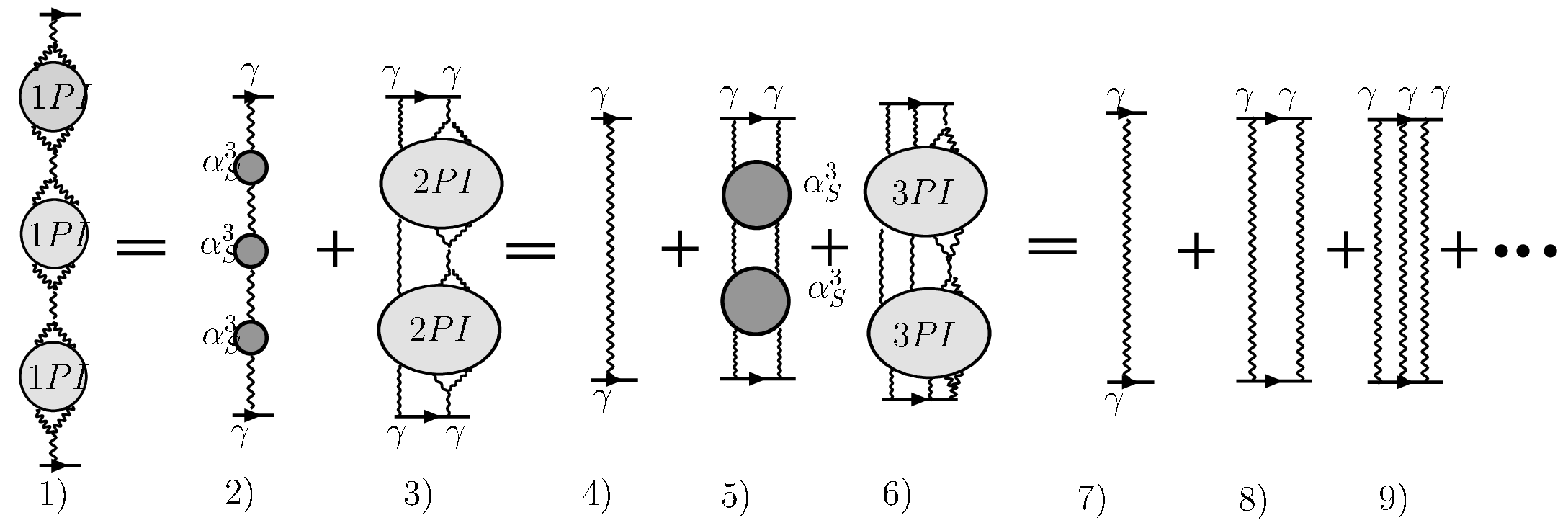,width=180mm,height=45mm}} \caption{ Reduction of the enhanced diagrams to  a  system of non-interecting Pomerons
. }
\label{dtrre0} }

The first singularity that  $\Sigma_1\Lb 1PI|\omega\Rb$   has is the pole $\omega - \omega(\gamma_1) + \omega(\gamma_2)$ . These poles are shown explicitly in \fig{dtrre0}-3 and closing contour on these poles we obtain the
contribution which is similar to \eq{R3}, namely,
\beq \label{R4}
 A\Lb \fig{dtrre0}-5 \Rb\,\,\propto\,\,\frac{1}{m_2!}N^2_2(\omega = \omega(\gamma_1) + \omega(\gamma_2)/\,\Lb Y\,\,+\,\,O\Lb 1/\as \Rb \Rb^{m_2}\,\,\Sigma_1\Lb 2PI|\omega = \omega(\gamma_1 + \omega(\gamma_2)\Rb
\eeq
where $m_2$ is the number of the two Pomeron poles\footnote{We assume that $\gamma_1$ and $\gamma_2$ are the same  for all two Pomerons poles. Actually, only sum of $\gamma_1 + \gamma_2$ preserves. Strictly speaking we need to consider all two Pomerons poles a bit different and take separately each of them. It is easy to see that in doing so we get the same result as in \eq{R4} after integration over relevant $\gamma$'s};  $N_2$ is the vertex of dipole-two Pomeron interaction (see \fig{red}-A) and  $\Sigma_1\Lb 2PI|\omega\Rb$
is the sum of two Pomeron irreducable diagrams. Since the first term in  $\Sigma_1\Lb 2PI|\omega\Rb$ gives
the contribution which is proportional to $\as^3$ (see \fig{red}-B) we can  neglect all contribution except
with $m_2 = 0$ reducing the diagram of \fig{dtrre0}-5 to two Pomerons exchange of \fig{dtrre0}-8.
To take the contribution from three Pomerons exchange in the diagram of \fig{dtrre0}-3 we need to consider the diagram
of \fig{dtrre0}-6 and close the contour in $\omega$ over the poles $\omega = \omega(\gamma_1) + \omega (\gamma_2) + \omega(\gamma_3)$.  Repeating the same procedure and taking into account that for the vertex of dipole-three Pomerons we have the relation shown in \fig{red}-C we obtain the diagram of \fig{dtrre0}-9. Continuing this procedure we
see that we reduce the general enhanced diagram to the system of non-interacting Pomerons.

\FIGURE[ht]{ \centerline{\epsfig{file=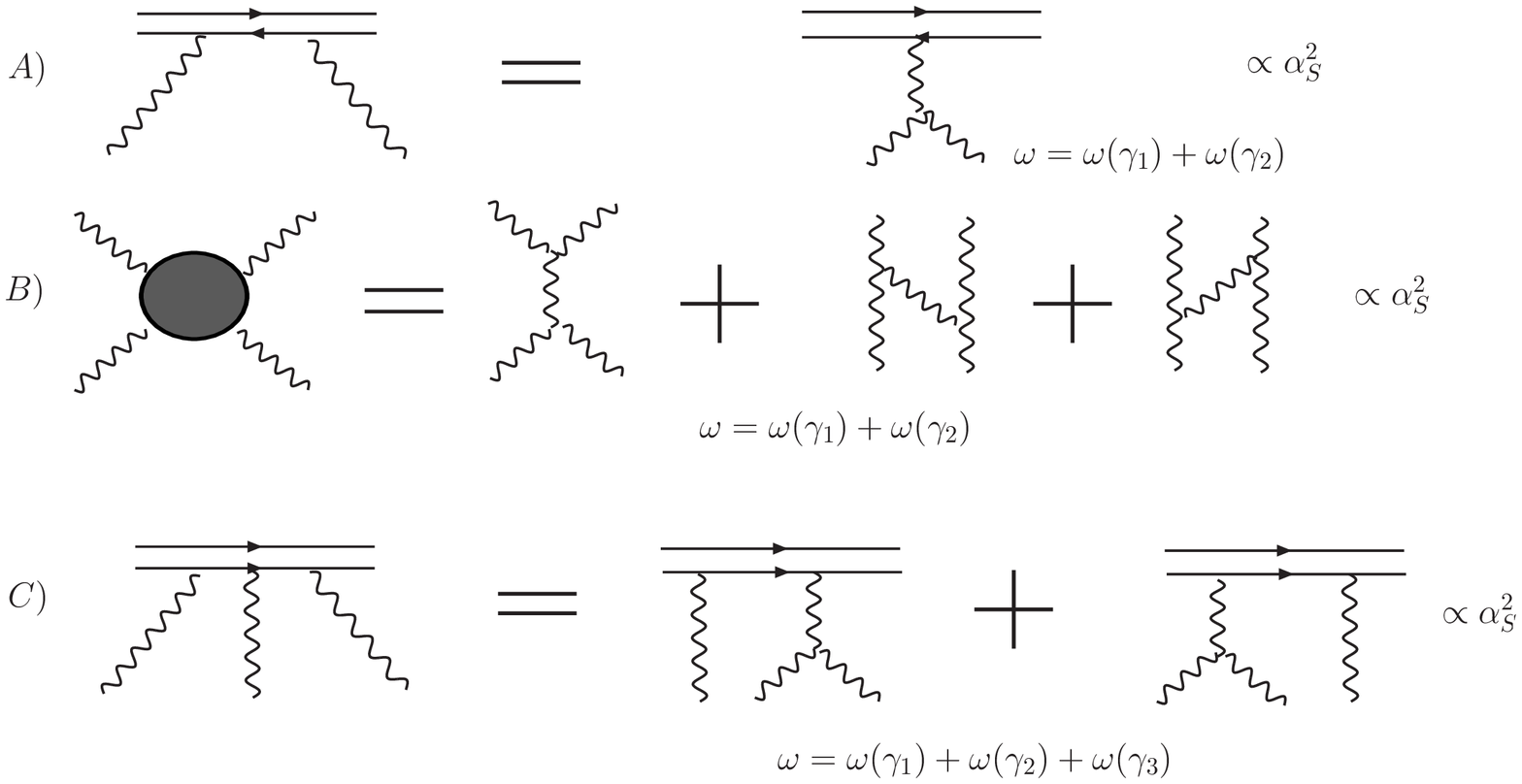,width=140mm}} \caption{ The vertices of two and three Pomerons interaction with dipoles and $\Sigma\Lb 2PI|\omega = \omega(\gamma_1) + \omega(\gamma_2)\Rb$
. }
\label{red} }
\section{ General solution for  the simplified BFKL kernel}
\subsection{The main idea}

The analysis of the first diagrams for the BFKL Pomeron calculus
shows that they have the same kind of contributions as for the BFKL
calculus in zero transverse dimensions and the new one that stems
from the possibility that for some specific values of the anomalous
dimension ($\gamma_{0,k}$) the intercept of  $n$  BFKL Pomerons in
exchange   is equal to the intercept of $k$     BFKL Pomerons  with
$k <n$.
 However, for this specific contribution two or more Pomerons are located inside the saturation region where we cannot use \eq{OM} to calculate the intercept of the Pomeron. Inside the saturation region  we can use \eq{OMSAT} which has a rather general
proof (see Ref. \cite{BALE}). In doing so we see that this specific
QCD contribution leads to an  energy suppressed
 scattering amplitude and can be neglected in the first approximation to the problem.

Having this result in mind we can restrict ourselves by calculating
only terms stemming from the decomposition
 of the type of \eq{OMI}.
This decomposition leads to the exchange of non-interacting BFKL
Pomerons and to the renormalisation of the vertex for the  Pomeron
target (projectile) interaction as well as to the  renormalisation
of the Pomeron intercept. In the kinematic region given by \eq{KR}
we can neglect the renormalisation of the Pomeron intercept and find
the renormalised  vertex for the Pomeron target (projectile)
interaction from the initial condition to our problem since it is
closely related to the dipole -dipole (or target) interaction at low
energies.

Since we are dealing with the system of non-interacting Pomerons we
can apply the procedure of the MPSI approximation (see Ref.
\cite{MPSI}) to take into account all the Pomeron loops. The
strategy of the improved MPSI approach consists of three major steps
\begin{enumerate}
\item \quad To find the solution in the  mean field approximation for arbitrary initial conditions.
According  to our analysis in this approximation the scattering
amplitude has the following form \beq \label{SA1} N\Lb Y-
Y',[\gamma_R(k_i,b_i)]\Rb \,\,=\,\,\sum_{n
=1}^{\infty}\,\int\,\prod^n_{i=1} \,d^2 k_i\,\,(-1)^n\,\,C_n(k)\,\,
P \Lb  k,k_i;b_i| Y -Y'\Rb\, \gamma_R(k_i,b_i)\,\, \eeq where
$\gamma_R$ is an arbitrary function of $k$ (recall that $k$ is the
conjugated variable to the size of a dipole).
\item \quad  Comparing the exact solution with the initial condition to find the solution in terms of the renormalized
vertices. It means that we need to solve the functional equation (in
the general case) \beq \label{SA2} N\Lb Y-
Y'=0,[\gamma_R(k,b)];b\Rb\,\,=\,\,N([\gamma(k,b)]) \eeq where the
function $N$ should be given. In other words the scattering
amplitude for low energies should be
 known either from calculations in QCD or from some phenomenology.
 This step is in full agreement with the parton model where the
  high energy interaction can be calculated through the low energy
   amplitude for the interaction of `wee' partons that have to be given.

\item \quad To use the MPSI formula to take into account the Pomeron loops.
This formula reads as follows \bea \label{IMPSIF} &&N^{IMPSI}\Lb
k,k_0;b,Y-0 \Rb \,\,= \,\,\sum_{n=1}^{\infty} \,\frac{(-1)^{n
-1}}{n!}\,\prod^n_{i=1,j=1}\,d^2 k_i d^2 k_j d^2 b_i d^2 b_j
\frac{\delta}{\delta\,\gamma^{(1)}_R(k_i,b_i)}\frac{\delta}{\delta\,\gamma^{(2)}_R(k_j,b_j)} \nonumber \\
&& \,N^{MFA}\Lb Y- Y',[\gamma^{(1)}_R(k_i,b_i)]\Rb\,N^{MFA}\Lb
Y',[\gamma^{(2)}_R(k_j,b_j)]\Rb|_{\gamma^{(1)}_R=\gamma^{(2)}_R
=0}\,\, \gamma^{BA}\Lb k_i, k_j ,\underline{b} - \underline{b}_i - \underline{b_k}\Rb
\eea where $\gamma^{BA}$ is the amplitude in the Born approximation.
It should be stressed that \eq{SA1} for $N^{MFA} $ guarantees that
$N^{IMPSI}\Lb k,k_0;b,Y-0 \Rb$ does not depend on the value of $Y'$.
Indeed, using the fact that \beq \label{SA3} \int\,d^2
k_i\,d^2\,k_j\,d^2b_i\,d^2 b_j  P \Lb  k,k_i;b_i| Y
-Y'\Rb\,\,\gamma^{BA}\Lb k_i, k_j ,\underline{b} - \underline{b}_i -
\underline{b_k}\Rb \,\,P \Lb  k_0,k_j;b_j| Y' \Rb\,\,\,=\,\,\bas^2\,P \Lb
k,k_0;b| Y \Rb \eeq one can rewrite \eq{IMPSIF} in the form \bea
\label{IMPSIF1}
&&N^{IMPSI}\Lb k,k_0;b,Y-0 \Rb \,\,=\sum_{n=1}^{\infty} \,\frac{(-1)^{n -1}}{n!}\,\prod^n_{i=1,j=1}\,d^2 k_i d^2 k_j \,d^2 b_i d^2 b_j \,\frac{\delta}{\delta\,\gamma^{(1)}_R(k_i,b_i)}\frac{\delta}{\delta\,\gamma^{(2)}_R(k_j,b_j)} \nonumber \\
&&\,N^{MFA}\Lb Y- Y',[\gamma^{(1)}_R(k_i,b_i)]\Rb\,N^{MFA}\Lb\,Y',[\gamma^{(2)}_R(k_j,b_j)]\Rb|_{\gamma^{(1)}_R=\gamma^{(2)}_R =0}\,\,\,\Lb \bar{\alpha}^2_S\,P \Lb  k,k_0;b| Y \Rb \Rb^n \nonumber \\
&&=\,\,\,1\,\,\,-\,\,\exp \left\{\,-\,\bar{\alpha}^2_S\,P \Lb
k,k_0;b| Y \Rb\,\frac{\partial}{\partial
\gamma^{(1)}_{P}}\,\frac{\partial}{\partial
\gamma^{(2)}_P}\,\right\}\,N^{MFA}\Lb \gamma^{(1)}_P \Rb\,
N^{MFA}\Lb  \gamma^{(2)}_P\Rb|_{\gamma^{(1)}_P\,=\,\gamma^{(2)}_P
\,=\,0} \eea where $\gamma^{(1)}_P\,\,=\,\,P \Lb  k,k_i;b_i| Y
-Y'\Rb\, \gamma_R(k_i,b_i)$ and $\gamma^{(2)}_P\,\,=\,\,P \Lb
k_0,k_i;b_i| Y'\Rb\,\gamma_R(k_i,b_i) $
\end{enumerate}

At the moment we can carry out this program only for a simplified
BFKL kernel since only for this kernel do
 we have the exact analytical solution to work with (see  \cite{LT}).
  The substantial number of numerical solutions \cite{NS} does not help us since we know how to perform
   the renormalisation of the vertices only for the analytical solution.

\subsection{ The simplified BFKL kernel and MFA  solution}
The kernel for which we find the solution and implement our program
has the following form \cite{LT}

\beq \label{CHIM} \omega(\gamma = \,\h +
i\nu)\,\,=\,\,\bas\,\,\left\{\begin{array}{c}
\,\,\,\,\,\,\frac{1}{\gamma}\,\,\hspace*{1cm}\mbox{for}\,\,r^2\,Q^2_s\,\ll\,1\,; \\
\,\frac{1}{1\,-\,\gamma}\,\,\hspace*{1cm}
\mbox{for}\,\,r^2\,Q^2_s\,\gg\,1\,;
\end{array} \right.
\eeq

This kernel  sums the leading log  contributions in the region of
low $x$. The first one is the usual LLA approximation of
perturbative QCD in which the terms
 of the order of
$ \left(\as\,\ln(r^2\,\Lambda^2)\right)^n$ are taken into account
for $r^2\,Q^2_s\,\ll\,1$ where
$\as\,\ln(r^2\,\Lambda^2)\,\,\gg\,\,1$; and the second log
approximation
 leads to the summation of the terms of the order
of $ \left(\as\,\ln(r^2\,Q^2_s)\right)^n$ in the kinematic region
 where $r^2\,Q^2_s\,\gg\,1$ and $\as\,\ln(r^2\,Q^2_s)
\,\gg\,1$ \cite{MULG,LT}.

The LLA leads to the ordering of the dipole momenta.

In the perturbative QCD region ($r^2\,Q^2_s\,\ll\,1$) we have, using
the diagrams of \fig{1endi} as an example, \beq \label{SK1} k
\,\gg\,k_1\gg \,k_2\,\gg \,k_0 \eeq

In the saturation region the ordering is  the opposite, namely,

\beq \label{SK2} k \,\ll\,k_1\ll \,k_2\,\ll\, \,\mbox{ max( $k_0$ or
$Q_s$)} \eeq where $Q_s$ is the saturation scale.

\eq{CHIM} is not only much simpler than the exact kernel  of \eq{OM}
but calculating the Pomeron diagrams in LLA we can neglect the
contribution due to the  overlapping of the intercept for a
different number of Pomerons in exchange. Indeed, the fact that a
number of Pomerons are in the saturation region while other Pomerons
are outside this region,  means that we cannot keep the ordering of
\eq{SK1} or \eq{SK2}. Therefore, dealing with this kernel we enhance
the arguments for neglecting the overlapping singularities.

Fortunately, the solution for the MFA has been found in
Ref.\cite{LT} in the entire kinematic region for arbitrary initial
conditions. For completeness of our presentation, we will describe
the main features
 of this solution inside of the saturation region in this section. The solution is written for the amplitude
 in the coordinate representation, namely
\beq \label{SK3} N\Lb x_{12};b;Y \Rb
\,=\,\,x^2_{12}\int\,\frac{d^2 k}{(2 \pi)^2}\,e^{i \underline{k} \cdot
\underline{x}_{12}}\,\,N\Lb k, b;Y\Rb \eeq

The solution in the MFA ( to the Balitsky-Kovchegov equation) can be
written in the form \beq \label{SK4} N\Lb z
\Rb\,\,=\,\,1\,\,-\,\,e^{-\zeta(z)} \eeq where the variable \beq
\label{ZV} z\,\,\,=\,\,\ln \Lb x^2_{12}\,Q^2_s(Y,b)\Rb \eeq and
function $-\zeta(z)$ is determined by the following equation \beq
\label{SK5}
z\,\,\,=\,\,\,\sqrt{2}\,\int^{\zeta}_{\zeta_0(b)}\,\frac{d
\zeta'}{\sqrt{\zeta'\,\,+\,\,( \exp \Lb - \zeta' \Rb \,-\,1)}} \eeq

The boundary conditions for  the solution of \eq{SK4} we should find
from the solution at $z <0$. We assume that to the right of the
critical line $
 x^2_{12}\,Q^2_s(Y,b)\,\approx\,1$ ($z = 0$) the exchange of one BFKL Pomeron gives the main contribution. It has the form \cite{IIM} for our simplified BFKL kernel
 \beq \label{SK6}
 P\Lb z, \gamma\Rb \,\,=\,\,\gamma\,e^{\h \,z}\,e^{ - b^2/R^2}\,\,=\,\,\gamma(b)\,e^{\h \,z}\,
 \eeq
 where $R^2$  ( in the spirit of the LLA approximation )  depends only on initial conditions at low energy and has a clear non-perturbative origin.
Therefore the boundary conditions have the form \bea
N\Lb z = 0_+ ;\eq{SK4}\Rb \,\,&=&\,\,N\Lb z = 0_-\Rb\,\,=\,\,\gamma\,\,e^{ - b^2/R^2} \,\,\equiv\,\gamma(b)\,; \label{BC1}\\
\frac{d \,\ln N}{ d z}|_{z = 0_+}\,&=&\,\,\frac{d \,\ln N}{ d z}|_{z
= 0_-}\,\,=\,\,\h\,; \label{BC2} \eea For small $\zeta$ the solution
to \eq{SK5} has the form \beq \label{SK7} \ln
\frac{\zeta}{\zeta_0(b)}\,\,=\h\,z\,\,\,\,\mbox{or}
\,\,\,\,\zeta\,\,=\,\,\zeta_0(b)\,e^{\h z} \eeq This equation says
that we can satisfy \eq{BC1} and \eq{BC2} if \beq \label{SK8}
\zeta_0(b)\,\,=\,\,\gamma(b) \,\,\,\,\mbox{and}
\,\,\,\,\,\gamma(b)\,\ll\,\,1 \eeq Since $\gamma(b) \,\approx
\,\bas^2$  \eq{SK8} shows that \eq{SK5} gives the solution to our
problem.

From \eq{SK5} one can see that at small values of $\zeta_0$ the
integral has a logarithmic divergence.  Therefore, we can rewrite
\eq{SK5} as follows \beq \label{SK9} z\,\,\,+2\, \ln \gamma(b)
=2\,\ln \Lb \gamma(b)\,e^{ \,\h
\,z}\Rb\,\,\,=\,\,\,\sqrt{2}\,\int^{\zeta}_{\zeta_0(b)}\,\frac{d
\zeta'}{\sqrt{\zeta'\,\,+\,\,( \exp \Lb - \zeta' \Rb \,-\,1)}} + \ln
\zeta_0(b) \eeq
 The r.h.s. of this equation does not depend on $\zeta_0(b) = \gamma(b)$ (see \eq{SK8}).
In \fig{zzeta} we plot $\zeta(z)$ for two cases  : the exact
expression of \eq{SK5} and \beq \label{SK10}
\sqrt{2}\,\int^{\zeta}_{a}\,\frac{d \zeta'}{\sqrt{\zeta'\,\,+\,\,(
\exp \Lb - \zeta' \Rb \,-\,1)}} + \ln \Lb \zeta_0(b)/a\Rb \eeq One
can see that both functions coincide. This is an illustration that
the integral of \eq{SK5} has no further dependence on $\gamma(b)$.
Having this in
 mind we can claim that \eq{SK9} leads to a general function, namely
\beq \label{SK12} \zeta(z)\,\,=\,\,\Phi\Lb \gamma(b) \,e^{\h \,z}
\Rb\,\,\equiv\,\,\Phi\Lb \gamma_P(z,b)\Rb \eeq The argument of the
function $\Phi$ is the one Pomeron exchange (see \eq{SK6}) and
expanding $ \Phi\Lb \gamma(b) \,e^{\h \,z} \Rb$ in a series with
respect to this argument we are able to find the coefficient $C_n$
in \eq{SA1}.

 \eq{SK5} has also two simple analytical solutions for the function $\Phi$: in the region of small $\zeta$ (
see \eq{SK7} and for large $\zeta$. The last solution is \beq
\label{SK11} \zeta\Lb z\Rb \,\,=\,\,\h\,\ln^2\Lb \gamma(b)\,e^{\h
\,z} \Rb \eeq The exact solution is shown in \fig{zetaz} together
with these two analytical solutions. One can see that these two
analytical solutions give a good approximation for small and large
values of $ \gamma(b)\,e^{\h \,z}$.
\DOUBLEFIGURE[ht]{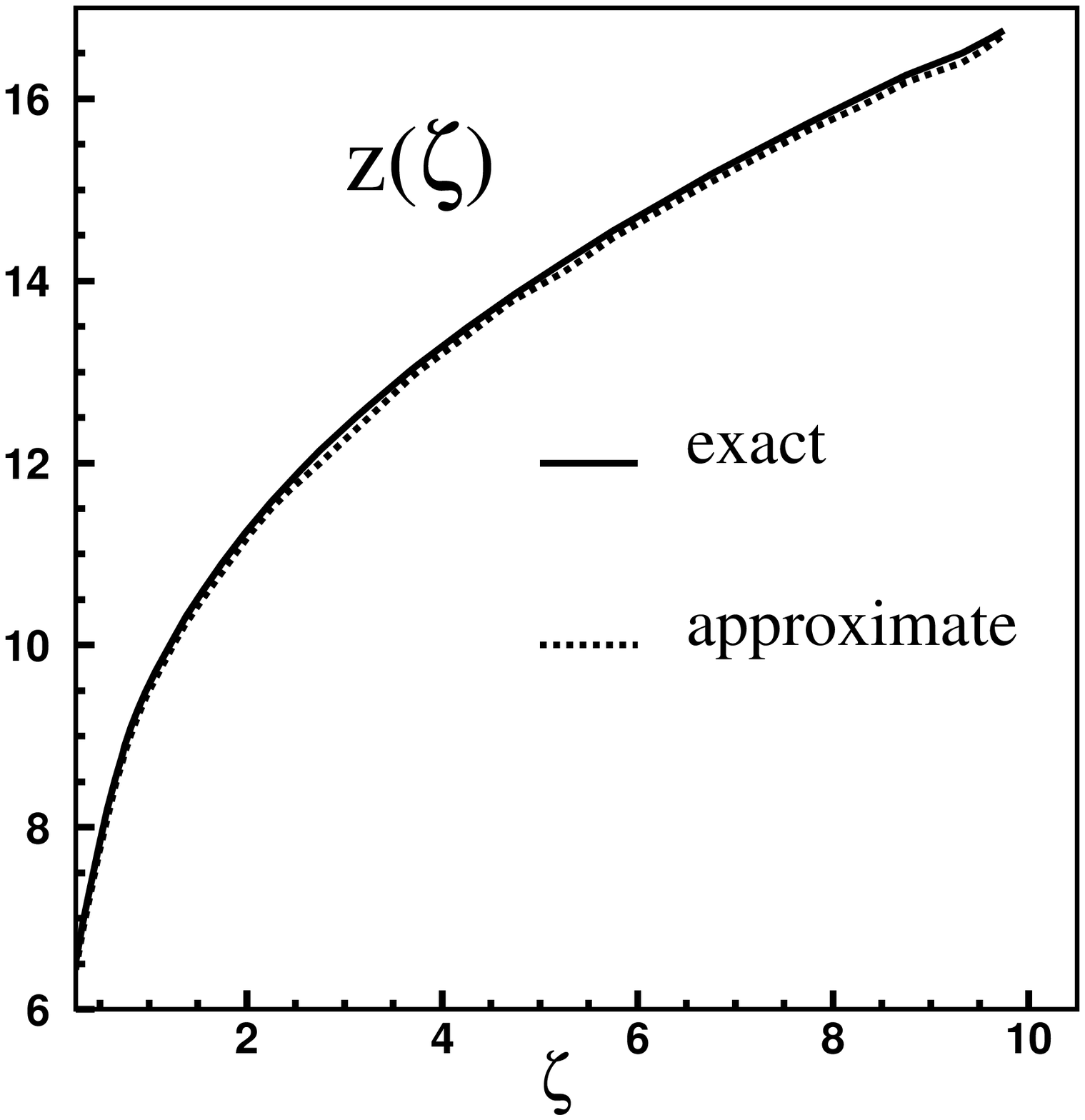,width=70mm,height=50mm}{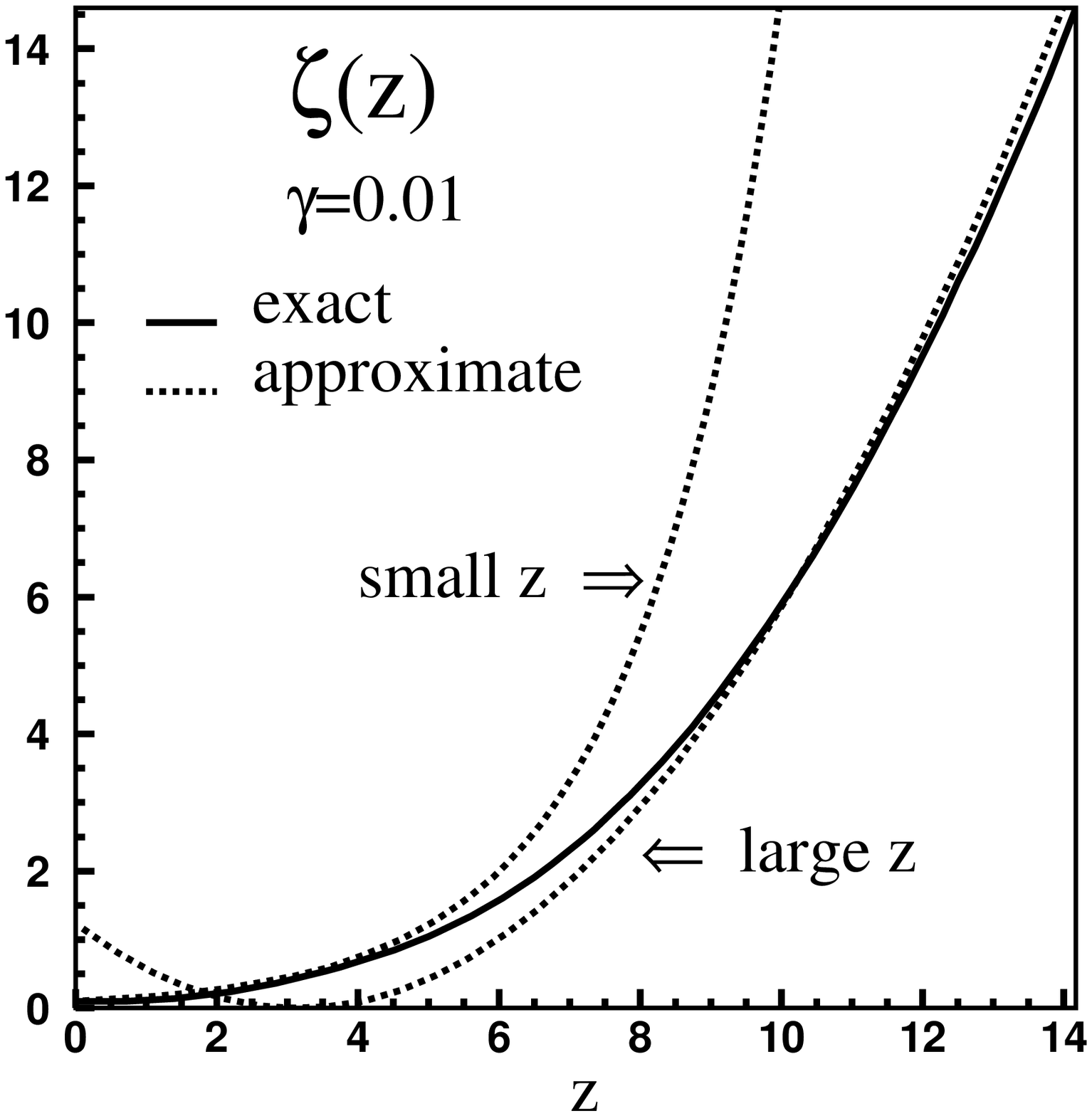,width=70mm,height=50mm}
{Comparison of the exact integral of \protect\eq{SK5} with the
approximate formula of \protect\eq{SK10} with $a = 0.25$.
\label{zzeta}}{Solution to \protect\eq{SK5} (solid line) and two
approximate analytical function: \protect\eq{SK7} (small $z$) and
\protect\eq{SK11} at large $z$ (both are shown in dotted lines).
\label{zetaz} }
\subsection{Improved MPSI solution for Pomeron loops}

Using \eq{SK4} and \eq{SK12} we can find the sum of the Pomeron
loops by means of \eq{IMPSIF1}. We demonstrate the answer by solving
analytically two extreme cases: (i) the behaviour of the scattering
amplitude near to the saturation scale;  and (ii)  the asymptotic
behaviour deep inside of the saturation region.

For the first case $\zeta = \gamma(b) \exp\Lb \h z\Rb$ and \eq{SK4}
together with \eq{IMPSIF1} lead to the following answer \beq
\label{PLO1} N^{IMPSI}\Lb x^2,R^2;Y \Rb\,\,=\,\,1 - \exp\Lb -
\gamma(b)\,P\Lb x^2;R^2;Y \Rb \Rb\,\,=\,\,1 - \exp\Lb -
\gamma(b)\,e^{\h z} \Rb \eeq
where $x$ and $R$ are the sizes of the projectile and target dipoles
and $z$ is defined by \eq{ZV} where $Q_s^2 \propto 1/R^2$.

For finding the behaviour of the scattering amplitude deeply inside
of the saturation region we will use \eq{PL3}.  In  new variables:
$\h l \, - \,l_1\,\,=\,\,\ln \Lb \gamma^{(1)}_P \Rb$ and
  $ \h l \, + \,l_1\,\,=\,\,\ln \Lb \gamma^{(2)}_P \Rb$, the contour $C$ looks as it is shown in \fig{c}.
  Since $N^{MFA}$ does not have singularities  in the  variables $l$ and $l_1$ we can replace the contour $C$ by
 the contour $C_R$ (see \fig{c}).
Therefore,  we can rewrite \eq{PL3} in the following form \bea
\label{PLO2} N^{MPSI}_0\Lb  z \Rb \, &=&\, 1 \,\,-\,\,
\frac{1}{(2\,\pi\,i)^2}\,\oint_{C_R}\!\!\!\!\!\!d \,l\,
\oint_{C_R}\!\!\!\!\!\! d\,l_1\,
 \,\,\exp \Lb \frac{e^{l }}{\,P\Lb\gamma^{BA}, z \Rb}\,\Rb\,\,\Lb \frac{ e^l}{\,P\Lb\gamma^{BA},z\Rb} \Rb
\Gamma\Lb 0,  \frac{ e^l}{\,P\Lb\gamma^{BA}, z \Rb}\Rb   \nonumber\\
 & \times &
 Z^{MFA}\Lb P\Lb e^{\h l \,+\,l_1} \Rb \Rb \, Z^{MFA}\Lb P\Lb e^{\h l - l_1}\Rb \Rb
\eea Using \eq{SK6} and \eq{SK11} we rewrite \bea \label{PLO3}
 Z^{MFA}\Lb P\Lb e^{\h l \, + \,l_1} \Rb \Rb\,\,&=&\,\,1 - N^{NFA}\Lb \gamma^{(1)}_P \Rb\,\,=\,\,e^{-\zeta(\h l + l_1)}\,=\,e^{ - \h \Lb \h l\,\, +\,\, l_1\Rb^2} \\
 Z^{MFA}\Lb P\Lb e^{ \h l \,-\, l_1} \Rb\Rb &=&\,\,1 - N^{NFA} \Lb\gamma^{(2)}_P   \Rb\,\,=\,\,e^{- \zeta(\h l - l_1)}\,\,=\,e^{ - \h \Lb \h l\,\, -\,\, l_1\Rb^2}\,\, \nonumber
\eea where $z = z_1 + z_2$.

One can see that the integration over $l_1$ reduces to a very simple
integral, namely \beq \frac{1}{2 \pi
i}\,\oint_{C_R}\!\!\!d\,l_1\,e^{-l^2_1}\,\,=\,\,\frac{1}{\pi}\,\,e^{\pi^2}\int^{\infty}_0\!\!d\,l_1\,sin\Lb
2\,\pi\,l_1 \Rb\,e^{-
l^2_1}\,\,=\,\,\frac{1}{2\,\sqrt{\pi}}\,erfi(\pi) \eeq where
$erfi(z) = erf(i\,z)/i$ and $erf(z)\,\,\equiv\,E_2(z)\,\,=\,\,\frac{2}{\sqrt{\pi}}\,\int^z_0,\exp\Lb - \,t^2 \Rb\,d t $
is the error function (see Refs. \cite{RY,AS}) .

The remaining integral over $l$ has the form \bea \label{PLO4}
&&N^{MPSI}_0\Lb  z \Rb \, \,\,= \,\,1 \\
&&\,- \,\frac{erfi(\pi)}{2
\sqrt{\pi}\,\pi}\,\,\int^{\infty}_0\!\!\!\!\!\!d \,l\,\left\{1 \,-
 \,\exp \Lb \frac{1}{\,\gamma^{BA} \,e^{\h z - l}}\,\Rb\,\,\Lb \frac{ 1}{\,\gamma^{BA}\,\,e^{\h z - l}} \Rb
\Gamma\Lb 0,  \frac{ 1}{\gamma^{BA}\,e^{\h z - l}}\Rb  \right\}
\,\times\,\sin\Lb 2 \pi l \Rb\,e^{ - \frac{1}{8}\,l^2} \nonumber
\eea It is clear from \eq{PLO4} that the dominant contribution stems
from $l $ of the order of unity due to the exponential decrease at
large $l$.  Therefore, the  behaviour of
 the integrand at large $\exp{( \h z)}$ will dictate the resulting approach to 1 for the amplitude.
 At $z \gg 1$ the integrand gives $\gamma(b)\,\exp{(- \h z + l)} $ behaviour. We can take
the integral over $l$ which leads to the following answer:
\bea
\label{PLO5}
 && \frac{erfi(\pi)}{2 \sqrt{\pi}\,\pi}\,\,\int^{\infty}_0\!\!\!\!\!\!d \,l\,\left\{ \gamma(b)\,e^{\h z - l} \right\} \,\times\,\sin\Lb  \pi l /2\Rb\,e^{ - \frac{1}{8}\,l^2}\,\,=\,\,\,\\
 &&\frac{erfi(\pi)}{2 \sqrt{\pi}\,\pi}\Lb i e^{2}\,\sqrt{\frac{\pi}{2}}\,\Lb
 erf\Lb \frac{4 + i \pi}{2 \sqrt{2}}\Rb \,\,+\,\,i \,\,erfi\Lb \frac{4 + i \pi}{2 \sqrt{2}}\Rb\Rb \Rb\,\,\frac{1}{\gamma(b)\,e^{  \h z}}\nonumber
 \eea
 The
solution of \eq{PLO5} shows the geometrical scaling behaviour and at
high energies $N \,\longrightarrow \,\,1\,-\,\exp\Lb - \h z \Rb$.
 In such a behaviour two features look unexpected: the geometrical scaling behaviour since the statistical
 physics motivated approach leads to a violation of this kind of behaviour \cite{STPH}; and the fact that the
 asymptotic behaviour shows a very slow fall down in comparison with
the expected $N \,\longrightarrow \,\,1\,-\,\exp\Lb - C\,z^2 \Rb$
\cite{LT, MPSI, KOLE}.
 This slow decrease shows that  the Pomeron loops essentially change the behaviour of
 the amplitude in the saturation region.

 \FIGURE[ht]{\begin{minipage}{80mm}
\centerline{\epsfig{file=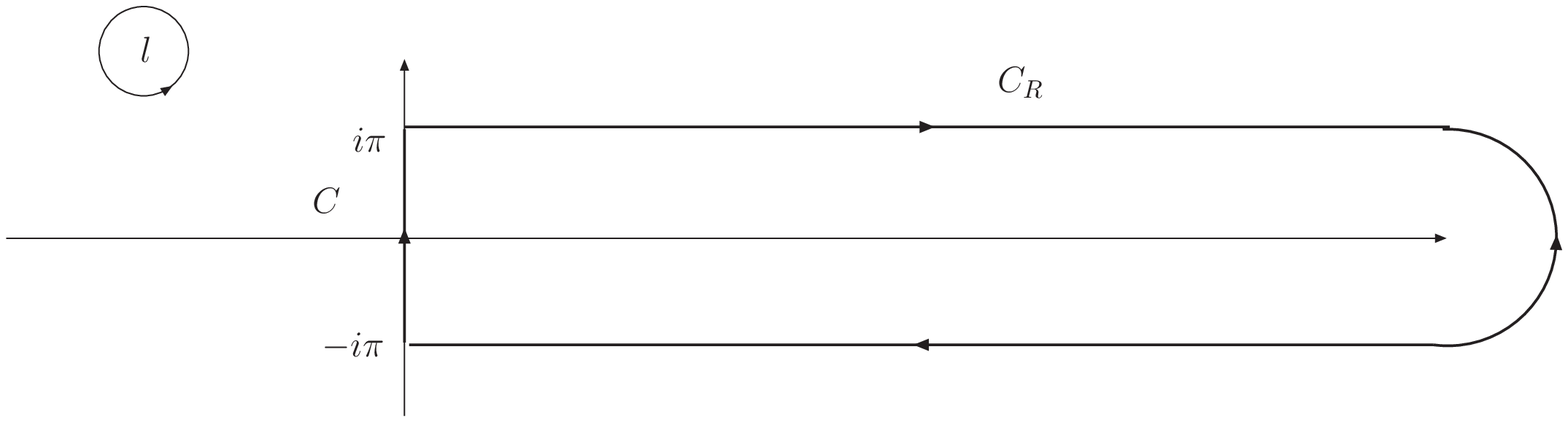,width=75mm}}
\end{minipage}
\caption{ The contours of integration in \protect\eq{PLO2}. }
\label{c} }

Our main difference to the statistical physics motivated approach is
the fact
 that we took into account the impact parameter behaviour of the scattering amplitude.
  We observed that the integration in the Pomeron loops is dominated by the impact
  parameters of the order of the largest dipole size involved in two vertices assigned to the loop.
  We can face two different possibilities. In the first case
  the  largest dipole belongs to the vertex , for which we assign  our rapidity  and for which the generating functional or other equation for our statistical system is written. This is the lower vertex of the Pomeron loop at the lower value of rapidity.
   In this case everything is in agreement with the statistical approach.
 In the second case the largest
    dipole belongs to the vertex  at the top of the Pomeron loop at higher  rapidity value which is our future in terms of the statistical approach.
     The future behavior of this situation cannot be described by a Markov
      chain without taking into account the long range rapidity correlations.
      It should be stressed that the appearance of the overlapping singularities cannot be included in the
        statistically  motivated formalism until the problem is reformulated in terms   of the interaction
        of new effective particles  related to these singularities.

\subsection{Self-consistency check}

As has been discussed we need to come back and consider the
overlapping singularities that have been neglected.
 The check of self-consistency of our approximation means that we need to calculate all diagrams but
replace the exchange of the BFKL Pomerons by the sum of all enhanced
diagrams (see \fig{ccendi-1}) in the kinematic region where the BFKL
Pomeron enters inside of the saturation region. For  example,
 the simplest diagram of \fig{1endi} has to be replaced by the diagram of \fig{ccendi-2} in
  the kinematic region that
is related to the overlapping singularities.

For the simplified kernel the value of $\gamma_0$ (see \fig{om0s})
is equal to $0.6$ and this value is close to $\gamma_{cr} = 1/2$.
However, since $\gamma_0 - \gamma_{cr}$ does not depend on the value
of rapidity, at high energies the solution that we need to check
will be at $z \,\gg\,1$.

 Indeed, the solution of  \eq{1P2} has the form of \eq{SK6} only at small values of $\nu$. In the region of small $\nu$ the expression of \eq{1P2} has the form
 \bea \label{CCC}
P\Lb  k,k_0|Y \Rb\,\,&=&\,\,\frac{1}{\sqrt{k^2 \,k^2_0}}\,\int\,d \nu\,e^{ \omega(\nu) Y}\,\Lb \frac{k^2}{k^2_0} \Rb^{i \nu} \nonumber \\
&\xrightarrow{\nu \,\ll\,1}&  \,\,\int\,d \nu\,\,\exp \Lb \h z\,- i \nu z\,\,\,-\,\,\,8\,\nu^2\,\bas Y \Rb \,\,\nonumber \\
 &\propto& \exp \Lb + \h z\,-\,\,\frac{z^2}{32\,\bas\,Y} \Rb
\eea In the last line of \eq{CCC} we used the steepest decent method
with the saddle point value of $\nu$
 \beq \label{SPNU}
 \nu
=\nu_{SP}\,\,=\,\, - i \frac{z}{16 \bas Y} \eeq
Therefore,  \eq{SK6}
for the Pomeron holds only for $z\,\ll\,4\,\sqrt{\bas Y}$ and at
$\nu_{SP} \ll\,1$. In this kinematic region $\nu_{SP} \,\ll\,1$. In
other words,  the region $\gamma \,=\,\h \,i \nu \to \gamma_0$
corresponds to the large values of $z$ where our scattering
amplitude behaves as
 \beq \label{CCC1}
A\,\,\,\longrightarrow\,\,\,1\,\,-\,\,e^{- \h z} \eeq

\DOUBLEFIGURE[ht]{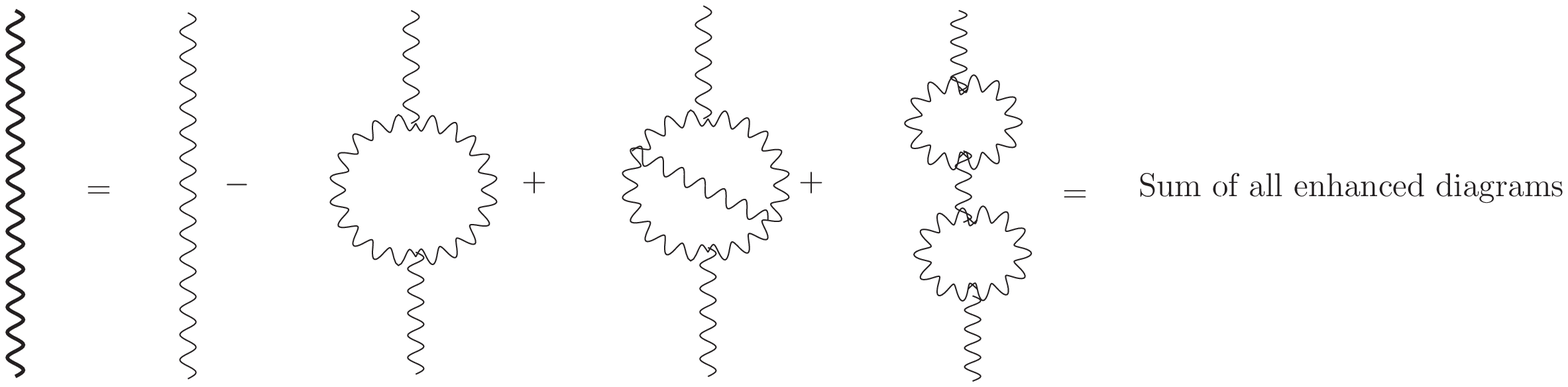,width=90mm,height=35mm}{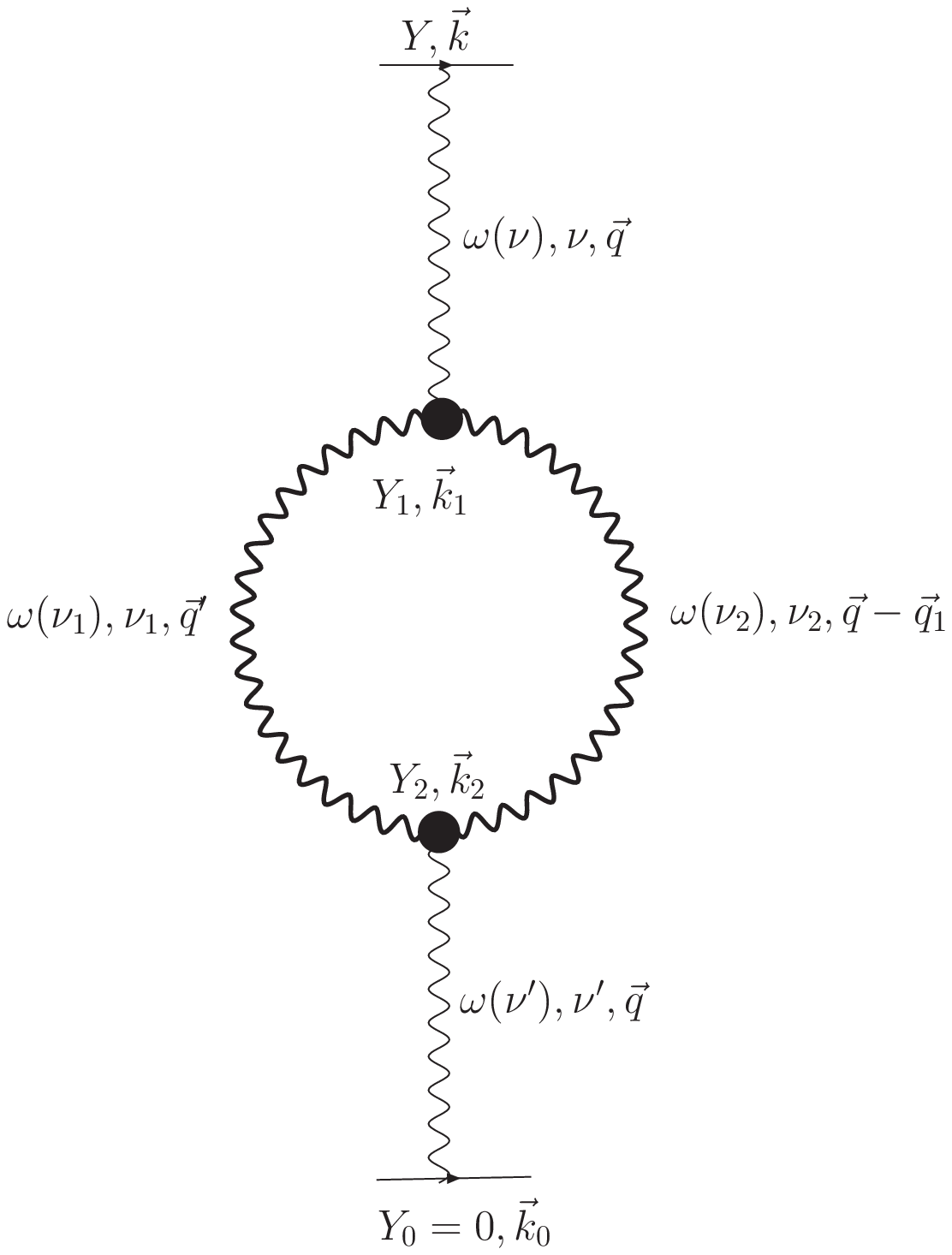,width=35mm,height=35mm}
{The graphic form for notation of wave bold line : the sum of all
enhanced diagrams.\label{ccendi-1}} { The diagram that we need to
replace the one in \protect\fig{1endi} to study the overlapping
singularities. By bold line we denote the sum of diagrams of \protect\fig{ccendi-1}. \label{ccendi-2}}

 For  the Pomeron near and inside of the saturation region we can rewrite the Pomeron diagrams in terms
 of the variables $z$ and $k$ introducing the Mellin transform for the Pomeron in the form
\beq \label{CC} P\Lb z \Rb \,\,=\,\,\,\frac{1}{2 \pi i}\,\,\int^{a +
i \infty}_{a - i \infty}\!\!\!\!\!\!d \lambda\,\,\frac{e^{\lambda
\,z}}{\lambda - \Delta} \eeq where the intercept    $\Delta_0 = \h$.

\eq{CCC1} has the form in $\lambda$-representation: \beq \label{CC2}
A\,\,=\,\,\frac{1}{\lambda} \,\,\,-\,\,\frac{C}{\lambda + \Delta_0}
\eeq the coefficient $C$ in \eq{CC2} reflects the fact that we know
the asymptotic behaviour
 inside the saturation region within the pre-exponential accuracy.

In this representation the diagram of \fig{ccendi-2} will be
proportional to \beq \label{CC1}
 A\Lb \fig{ccendi-2}
\Rb\,\,\propto\,\,\frac{1}{2 \pi i}\,\,\int^{a + i \infty}_{a - i
\infty}\!\!\!d \lambda\,\,\, e^{\lambda z}\,\,\,\frac{1}{\lambda -
\Delta_0}\,\,\,\left\{ \frac{1}{\lambda} \,\,\,-\,\,\frac{C}{\lambda
+ \Delta_0}   \right\}\,\,\,\frac{1}{\lambda - \Delta_0} \eeq
instead of \eq{FED}.

 One can see that we do not have the overlapping singularities. The main contribution stems from $\lambda =
 \Delta_0 $ and, since we have a double pole  for this value of $\lambda $ the diagram describes the renormalisation
  of the Pomeron intercept which is proportional to $\bas^3$ and can be neglected in the kinematic region of \eq{KR}.

\section{Conclusions}

The main results of this paper are:
\begin{enumerate}
\item \quad In the kinematic region of \eq{KR} the BFKL Pomeron calculus after renormalisation of the Pomeron
- target vertex,
 can be reduced to summing over the non-enhanced diagrams that describe the system of non-interacting Pomerons;
\item \quad The sum of enhanced diagrams for the system of non-interacting Pomerons can be calculated by
 means of  the improved Mueller-Patel-Salam-Iancu approach, using an additional piece of information on dipole
 scattering amplitude at low energy (for example, using the Born Approximation of perturbative QCD for
 its determination);
\item \quad For the simplified BFKL kernel given by \eq{CHIM} the analytical solution was found
to have  two unexpected features: the geometrical scaling behaviour
is deeply in the saturation region
 and is slow to  approach
 ($N \,\to 1 - C\exp( - \h z)$ instead of $ N \to 1 - C\exp(- z^2/8)$ as it happens in the MFA formalism.
\end{enumerate}

In this  paper we confirm the result of Hatta and Mueller \cite{HM}
that a new type of singularities appear in
 the BFKL Pomeron calculus: the overlapping singularities. They lead to a  steeper asymptotic behaviour than
 the multi BFKL Pomeron exchanges give. It means that even the MFA  formalism is doubtful without talking about
 summing over Pomeron loops. We argue that we can neglect these singularities since they stem from the kinematic
 region where two or more BFKL Pomerons are in the saturation region. We did the self-consistency check that such
  kind of singularities do not appear in the framework of the solution that we found.

As has been mentioned several times in this paper,  the geometrical
scaling behaviour is in contradiction with the statistical physics
analogy description and the Langevin equation based on this
analogy\cite{STPH}.  However, we need to recall that the
statistical-like approach at the moment exists only as a QCD
motivated model since the form of the noise term that has been used,
is oversimplified. Indeed, we can prove that the BFKL
 Pomeron calculus can be described by the Langevin equation with some noise term but this noise is  so  complicated
( see  \cite{KLPP})  that we do not know how to treat it. Instead of
this the simplified model for the noise term
 has been used. This model, as far as we understand it, has its own difficulties: (i) it cannot describe the overlapping
  singularities that have been discussed here;
  (ii) it did not include the impact parameter dependence which is very important for any treatment of the
  Pomeron loops as it was shown here; and (iii) even for the toy model in which we neglect the dependence of the  vertices
   on the dipole sizes, the statistical model predicts the
constant cross section at high energies \footnote{ We got this
information from S. Munier (lecture at GGI workshop on high density
QCD) and from E. Naftali (private communication).}
 while the exact solution (see Ref. \cite{TM} ) leads to a decrease of the total cross section.

We believe that the most vital problem, among these, is the correct
impact parametre dependence. The calculation of the enhanced
diagrams show that the typical impact parameter in the Pomeron loops
is of the order of the smallest size of the dipoles involved in the
two vertices in the loops. It might
 or might not  be the dipole from the vertex at rapidity Y for which we write the equation. If this
 dipole belongs to the second vertex, it is related to how we proceed from here and  we do not think that it is possible
  to describe such a system in a statistical way, assuming only short range correlations in rapidity.
The failure of the toy model description by the Langevin equation is
, in our opinion, due to the fact
 that a statistical model cannot describe the renormalization of the intercept of the Pomeron and their
 applicability is restricted to the kinematic region of \eq{KR}.

Since we consider the case of fixed coupling , even in deep inelastic scttering we have to take into acount all Pomeron loops. Indeed, as it was shown in Ref. \cite{GLR} only running QCD coupling selects the 'fan' diagrams and leads to the mean field approximation  for these processes. One cn see that the sum of Pomeron loopsat $Q^2$ near to the saturation scale has a typical eikonal form (see \eq{PLO1}). This result supports the  numerous  saturation models that describes well the experimental data \cite{MOD}.

However, it seems more realistically, for  a description of the deep inelastic processes first to solve the problem of summation of the Pomeron loops in the case of running QCD coupling. In this case we expect \cite{GLR} the the mean field approximation will be able to describe the sunstantial part of kinematic region.
More that ten years ago  Braun and Levin\cite{RUNA} suggested the procedure how to include the running  QCD coupling in to th BKL equayion that describes the linear evolution in the region of high energy. This suggestion was based on the strong assumption (which by the way, were proved ) that the gluon reggeization being the essential part of the BFKL approach in the leading order will be preserved in the next-to-leading order as well. The formula that they obtained has a so called triumvirate structure, namely, for the basic process
$ G(\vec{q}) + G(\vec{q}') \to G( \vec{q} - \vec{q}' $) the  fixed coupling constand should be replaced by
$$
\,\,\as \,\to \frac{\as(\vec{q}  - \vec{q}')\cdot\as(\vec{q}')}{\as(\vec{q})}$$
Recently this conjucture has been proven (see Ref. \cite{TRI}. This breakthrough allows us to include the running QCD coupling in the equation of the high density QCD. However, an additional problem arise that  should be studied, namely, such way of taking into account the running coupling effect induces the contriutions of infrared  renormalons (see paper of Levin in Ref. \cite{RUNA} ).
Such a contribution cannot be treated perturbatively and can restrict the theoretical accuracy  of our approach.   Therefore, summing Pomeron loops with running QCD coupling is our next challenging problem.

\section*{Acknowledgments}
We want to thank Asher Gotsman, Alex Kovner,  Uri Maor, Larry
McLerran  and
 Al Mueller for very useful
discussions on the subject of this paper. Special thanks from A.P. goes to Carlos Pajares for his
hospitality and support in the University of Santiago de Compostela where this work was completed.

 This research was supported in part  by the Israel Science Foundation,
founded by the Israeli Academy of Science and Humanities, by a
grant from Ministry of Science, Culture \& Sport, Israel \& the
Russian  Foundation for Basic research of  the Russian Federation,
and by BSF grant \# 20004019.

We thank the Galileo Galilei Institute for Theoretical Physics for
the hospitality and the INFN for partial support during the
completion of this work.

\renewcommand{\theequation}{A-\arabic{equation}}
\setcounter{equation}{0}  

\begin{boldmath}
\section*{Appendix A: Calculation of $g\Lb\underline{k},\underline{q},n,\nu\Rb$.} \label{sec:A}
\end{boldmath}

We want to calculate the Fourier transform of the vertex function
$E$ defined in \eq{E-fourier} as

 \bea \label{f-1}
g(\underline{k},\underline{q},n,\nu)\equiv \int \frac{d^2
\underline{x}_{12}}{\underline{x}_{12}^2}\; d^2 \underline{z} \;e^{i
\underline{k}\cdot \underline{x}_{12}} \; e^{i \underline{q}\cdot
\underline{z}} \;
E(\underline{x}_1,\underline{x}_2;\underline{x}_0|n;\nu) \eea
with $\underline{z}=(\underline{x}_{1}+\underline{x}_{2})/2-\underline{x}_{0}$.
The similar Fourier transform was already calculated in \cite{Bartels:2001hw}, but the authors used a
different normalization and we found it useful to calculate it using our notation.
The integration over $\underline{z}$ was done by Navelet and Peschanski in \cite{NP} by considering the
solutions of the differential equations obeyed by the vertex functions $E$ and consequent
    matching of the result to an approximate one obtained by Lipatov in \cite{LI}.
We adopt the notation introduced by Lipatov for the mixed representation of the vertex functions
\bea \label{f-mixed}
E^{n,\nu}_q(\rho)=\frac{2\pi^2}{b_{n,\nu}}\frac{1}{|\rho|}\int dz d \bar{z}e^{\frac{i}{2}(\bar{q}z+q\bar{z})}
E^{n,\nu}(z+\rho/2,z-\rho/2)
\eea
where
\bea \label{f-b}
b_{n,\nu}=\frac{2^{4i\nu}\pi^3}{|n|/2-i\nu}
\frac{\Gamma(|n|/2-i\nu+1/2)\Gamma(|n|/2+i\nu)}{\Gamma(|n|/2+i\nu+1/2)\Gamma(|n|/2-i\nu)}
\eea
 and $\rho=x_{12} $ in our notation introduced in \eq{E-def}.
The expression for $E^{n,\nu}_q(\rho)$ is given in \cite{NP} and reads
\bea \label{f-e}
E^{n,\nu}_q(\rho)&=&\bar{q}^{\;i\nu-n/2}q^{\;i\nu+n/2}2^{-6i\nu} \Gamma(1-i\nu+|n|/2)
\Gamma(1-i\nu-|n|/2)\times \nonumber \\
&&
\left[J_{n/2-i\nu}(\frac{\bar{q}\rho}{4})J_{-n/2-i\nu}(\frac{q\bar{\rho}}{4})
-(-1)^n J_{-n/2+i\nu}(\frac{\bar{q}\rho}{4})
J_{n/2+i\nu}(\frac{\bar{q}\rho}{4})
\right]
\eea
Thus in terms of $E^{n,\nu}_q(\rho)$ the Fourier transform in \eq{f-1} is given by

 \bea \label{f-2}
g(\underline{k},\underline{q},n,\nu)=
\frac{b_{n,\nu}}{2\pi^2}
\int \frac{d^2
\underline{\rho}}{|\underline{\rho}|} \;e^{i
\underline{\rho}\cdot \underline{k}} \;
E^{n,\nu}_q(\rho)
\eea
From \eq{f-e} and \eq{f-2} one can see that the calculation of $g(\underline{k},\underline{q},n,\nu)$
is reduced to evaluation of  the following integral
\bea\label{f-i}
\mathcal{I}_{\mu,\mu'}(q,k)\equiv
\int \frac{d^2
\underline{\rho}}{|\underline{\rho}|} \;e^{i
\underline{\rho}\cdot \underline{k}} \;
J_{\mu}(\frac{\bar{q}\rho}{4})J_{\mu^\prime}(\frac{q\bar{\rho}}{4})
\eea
We introduce polar coordinates for complex variables
\bea \label{f-polar}
q=|q|e^{i \psi} \hspace{2cm}  \rho=|\rho|e^{i \phi} \hspace{2cm}  k=|k|e^{i \theta}
\eea
and rewrite \eq{f-i}  as
\bea\label{f-i-2}
\mathcal{I}_{\mu,\mu'}(q,k)\equiv
\int_0^{\infty} \frac{\rho \;d
\rho}{\rho} \int_0^{2\pi} d \phi\;e^{i
\rho k \cos(\phi -\theta)} \;
J_{\mu}(\frac{q\rho }{4}e^{i(\phi-\psi)})J_{\mu^\prime}(\frac{q\rho}{4}e^{-i(\phi-\psi)})
\eea
For the sake of simplicity in \eq{f-i-2} and further below we denote by $\rho$, $q$ and $k$
their respective absolute values.
We use the series representation for the Bessel functions
\bea \label{f-series}
J_{\kappa}(z)=\left(\frac{z}{2}\right)^\kappa \sum^{\infty}_{m=0}
\frac{(-1)^m }{m! \;\Gamma(\kappa+m+1)}\left(\frac{z}{2}\right)^{2m}
\eea
to recast \eq{f-i-2} into
\bea\label{f-i-3}
\mathcal{I}_{\mu,\mu'}(q,k)&=&
\int_0^{\infty} d\rho  \int_0^{2\pi}\;d\phi \;e^{i\rho k \cos(\phi-\theta)} \nonumber\times \\
&&e^{i\mu (\phi-\psi)} \left(\frac{q\rho}{8}\right)^\mu\sum^{\infty}_{m=0}
\frac{(-1)^m }{m! \;\Gamma(\mu+m+1)}\left(\frac{q\rho}{8}\right)^{2m}e^{i2m (\phi-\psi)} \times \\
&&e^{-i\mu'(\phi-\psi)} \left(\frac{q\rho}{8}\right)^{\mu'}\sum^{\infty}_{{m'}=0}
\frac{(-1)^{m'} }{{m'}! \;\Gamma(\mu'+m'+1)}\left(\frac{q\rho}{8}\right)^{2{m'}} e^{-i2{m'} (\phi-\psi)}\nonumber
\eea
Changing variables $\phi-\theta  \; \rightarrow \; \phi$ and using the integral representation of the Bessel functions
\bea \label{f-bessel-int}
J_{n}(z)=&&\frac{1}{2\pi}\int_{-\pi}^{\pi} e^{-i n \theta  + i z \sin\theta} d\theta=
\frac{1}{2\pi}\int_{-\pi-\pi/2}^{\pi-\pi/2} e^{-i n (\theta-\pi/2)  + i z \sin(\theta-\pi/2)}
d(\theta-\pi/2) \\
&&=
\frac{(-i)^n}{2\pi}\int_{-3\pi/2}^{\pi/2} e^{-i n \theta  + i z \cos\theta}
d\theta=\frac{(-i)^n}{2\pi}\int_{0}^{2\pi} e^{-i n \theta  + i z \cos\theta}
d\theta
\nonumber
\eea
we rewrite \eq{f-i-3} as
\bea\label{f-i-4}
\mathcal{I}_{\mu,\mu'}(q,k)=
\sum^{\infty}_{m=0} \sum^{\infty}_{{m'}=0} \int_0^{\infty} & d\rho&
 (2\pi)\;J_{-\mu-2m+ \mu' + 2m}(\rho k)
\times \nonumber\\
 &&i^{-\mu}e^{i\mu (\theta-\psi)} \left(\frac{q\rho}{8}\right)^\mu
\frac{1 }{m! \;\Gamma(\mu+m+1)}\left(\frac{q\rho}{8}\right)^{2m}e^{i2m (\theta-\psi)} \times \\
&&i^{\mu'}e^{-i\mu'(\theta-\psi)} \left(\frac{q\rho}{8}\right)^{\mu'}
\frac{1 }{{m'}! \;\Gamma(\mu'+m'+1)}\left(\frac{q\rho}{8}\right)^{2{m'}} e^{-i2{m'} (\theta-\psi)}
\nonumber
\eea
Now we perform the integration over $\rho$ using the formula
\bea \label{f-r-int}
\int_0^{\infty}d\rho \;J_{\alpha}(\rho k) \left(\frac{\rho q}{8}\right)^\beta
=\frac{1}{k}\left(\frac{ q}{4 k}\right)^\beta
\frac{\Gamma(\frac{1}{2}+\frac{\alpha+\beta}{2})}{\Gamma(\frac{1}{2}+\frac{\alpha-\beta}{2})}
\eea
then the expression in \eq{f-i-4} reads
\bea\label{f-i-5}
\mathcal{I}_{\mu,\mu'}(q,k)=
\frac{2\pi}{k} &&i^{-\mu}e^{i\mu (\theta-\psi)} \left(\frac{q}{4k}\right)^\mu\sum^{\infty}_{m=0}
\frac{1 }{m! \;\Gamma(\mu+m+1)}\left(\frac{q}{4k}\right)^{2m}e^{i 2m (\theta-\psi)}
\frac{1}{\Gamma(1/2-\mu-2m)}
 \times \\
&&i^{\mu'}e^{-i\mu'(\theta-\psi)} \left(\frac{q}{4k}\right)^{\mu'}\sum^{\infty}_{{m'}=0}
\frac{1 }{{m'}! \;\Gamma(\mu'+m'+1)}\left(\frac{q}{4k}\right)^{2{m'}} e^{-i2{m'} (\theta-\psi)}
\Gamma(1/2+\mu'+2m')
\nonumber
\eea
One should note the full factorization of two series in \eq{f-i-5} due to the integration over $\rho$.
This allows us to sum them separately using
\bea \label{f-sums-1}
\sum^{\infty}_{{m}=0}
\frac{z^{2{m}} }{{m}! \;\Gamma(\mu+m+1)}  \frac{1}{\Gamma(1/2-\mu-2m)}=
\frac{\cos(\pi\mu)}{\pi}
\frac{\Gamma(1/2+\mu)}{\Gamma(1+\mu)} \;_2F_1[1/4+\mu/2,3/4+\mu/2;1+\mu;4\;z^2] \hspace{2cm}
\eea
and
\bea \label{f-sums-2}
\sum^{\infty}_{{m}=0}
\frac{z^{2{m}} }{{m}! \;\Gamma(\mu'+m+1)} \Gamma(1/2+\mu'+2m)=
\frac{\Gamma(1/2+\mu')}{\Gamma(1+\mu')} \;_2F_1[1/4+\mu'/2,3/4+\mu'/2;1+\mu';4\;z^2]
\eea
With help of \eq{f-sums-1} and \eq{f-sums-2} we obtain the final expression for $\mathcal{I}_{\mu,\mu'}(q,k)$ as
\bea\label{f-i-6}
\mathcal{I}_{\mu,\mu'}(q,k)=
\frac{2\pi}{k} &&i^{-\mu}e^{i\mu (\theta-\psi)} \left(\frac{q}{4k}\right)^\mu
\frac{\cos(\pi\mu)}{\pi}
\frac{\Gamma(1/2+\mu)}{\Gamma(1+\mu)}
\;_2F_1[1/4+\mu/2,3/4+\mu/2;1+\mu;4\left(\frac{q}{4k}\right)^2e^{i 2(\theta-\psi)}]
 \times\nonumber \\
&&i^{\mu'}e^{-i\mu'(\theta-\psi)} \left(\frac{q}{4k}\right)^{\mu'}
\frac{\Gamma(1/2+\mu')}{\Gamma(1+\mu')}
 \;_2F_1[1/4+\mu'/2,3/4+\mu'/2;1+\mu';4\;\left(\frac{q}{4k}\right)^2e^{-i 2(\theta-\psi)}]
\hspace{1cm}
\eea
At this point we return to complex vector notation given in \eq{f-polar} and absorb the angles into
complex vectors $q$, $k$, $\bar{q}$ and $\bar{k}$
\bea\label{f-i-7}
\mathcal{I}_{\mu,\mu'}(q,k)=
\frac{2\pi}{|k|} &&i^{-\mu} \left(\frac{\bar{q}}{4\bar{k}}\right)^\mu
\frac{1}{\Gamma(1+\mu)\Gamma(1/2-\mu)}
\;_2F_1[1/4+\mu/2,3/4+\mu/2;1+\mu;\left(\frac{\bar{q}}{2\bar{k}}\right)^2]
 \times \hspace{1cm}\\
&&i^{\mu'}\left(\frac{q}{4k}\right)^{\mu'}
\frac{\Gamma(1/2+\mu')}{\Gamma(1+\mu')}
 \;_2F_1[1/4+\mu'/2,3/4+\mu'/2;1+\mu';\left(\frac{q}{2k}\right)^2]
\nonumber
\eea
In the first line in  \eq{f-i-7} we used identity $\Gamma(1/2+z)\Gamma(1/2-z)=\cos(z\pi)/\pi$.
Plugging \eq{f-i-7} into \eq{f-2} we find the Fourier transform of the vertex functions $E$
defined in \eq{f-1}
 \bea \label{f-full}
g(\underline{k},\underline{q},n,\nu)&=&
\frac{b_{n,\nu}}{2\pi^2}\;i^n \;\bar{q}^{\;i\nu-n/2}q^{\;i\nu+n/2}2^{-6i\nu} \Gamma(1-i\nu+|n|/2)
\Gamma(1-i\nu-|n|/2)
\times \nonumber \\
&&
\left[\mathcal{I}_{n/2-i\nu,-n/2-i\nu}(q,k)
-(-1)^n\mathcal{I}_{-n/2+i\nu,n/2+i\nu}(q,k)
\right]
\eea
with $b_{n,\nu}$ given in \eq{f-b}.
In the high energy limit the main contribution to Pomeron Green function comes from $n=0$.
We are interested in the case of small momentum transferred  along the Pomeron which implies
\beq \label{f-limit}
|k| \gg |q|
\eeq
In the limit of \eq{f-limit} for $n=0$ \eq{f-full} reads
\bea \label{f-full-2}
g(\underline{k},\underline{q},0,\nu)&=&
\frac{b_{0,\nu}}{2\pi^2} \;|q|^{\;2i\nu}2^{-6i\nu} \Gamma^2(1-i\nu)\cos(i\nu\pi)
\left[\mathcal{I}_{-i\nu,-i\nu}(q,k)
-\mathcal{I}_{+i\nu,+i\nu}(q,k)
\right]=
 \\
&& \simeq \frac{\pi^2}{-i\nu}2^{2i\nu}|k|^{-1+2i\nu}\frac{\Gamma^2(1/2-i\nu)\Gamma(i\nu)}{\Gamma^2(1/2+i\nu)\Gamma(-i\nu)}
-
\frac{\pi^2}{-i\nu}2^{-6i\nu}|k|^{-1-2i\nu}|q|^{4i\nu}\frac{\Gamma^2(1-i\nu)\Gamma(i\nu)}{\Gamma^2(1+i\nu)\Gamma(-i\nu)} \nonumber
\eea

\renewcommand{\theequation}{B-\arabic{equation}}
\setcounter{equation}{0}  

\begin{boldmath}
\section*{ Appendix B:  Triple Pomeron vertex in $q$ and $\nu$ representation.}\label{sec:B}
\end{boldmath}
In this appendix  we calculate the triple Pomeron vertex in $q$ and $\nu$ representation,  namely,
\beq  \label{B1}
\Gamma_{3P} \Lb q'_1, q'_2; \nu_1,\nu_2,\nu_3 \Rb
 \equiv \int\,d^2 k'_1\,\,\,g\Lb \underline{k}'_1,\underline{q}'_1,0,\nu_1 \Rb \,\,g^*\Lb  \underline{k}'_1 + \frac{1}{2} \underline{q}'_2, \underline{q'}_2 ,0, \nu_2\Rb \,\,
g^*\Lb  \underline{k}'_1 -\underline{q}'_1 + \frac{1}{2}  \underline{q}'_2,\underline{q}'_1- \underline{q}'_2 ,0,\nu_3\Rb
\eeq
where $g$ is given by \eq{f-full-2}.

Introducing new holomorphic variables
\bea \label{NV}
&&
t\,\,=\,\,\frac{q'_2}{2 k'_1 + q'_2}\,\; \,\,\,\,\mbox{and}\,\,\,\,\,\,
 \bar{t}\,\,=\,\,\frac{\bar{q}'_2}{2\,\bar{k}'_1  + \bar{q}'_2}
\eea
and using \eq{f-i-7} , we can rewrite \eq{B1} in the form
\bea
&&\Gamma_{3P} \Lb 0,q'_2; \nu,\nu_2,\nu_3 \Rb
=\,\,4\, \Lb \frac{{q'_2}^2}{4}\Rb^{1/2 - i \nu_1 + i \nu_2 + i \nu_3}\,\,\int d t d \bar{t}
\,\,\,\,\Lb ( 1 - t )\,( 1 - \bar{t})\Rb^{1/2 + i \nu_1}
 C_1(\nu_1)\label{B2} \\
&&\times \left\{ C^*_1(\nu_2) {}_2F_1\Lb 1/4 + i \nu_2/2, 3/4 + i \nu_2/2, 1 +i \nu_2, t^2 \Rb\,\,{}_2F_1\Lb 1/4 + i \nu_2/2, 3/4 + i \nu_2/2, 1 +i \nu_2, \bar{t}^2\Rb \right. \nonumber \\
&& \left.   - \,\,C^*_2(\nu_2)\Lb 4\,t\,\bar{t}\Rb^{2 i \nu_2}\, \,{}_2F_1\Lb 1/4 + i \nu_2/2, 3/4 + i \nu_2/2, 1 +i \nu_2, t^2 \Rb\,\,{}_2F_1\Lb 1/4 + i \nu_2/2, 3/4 + i \nu_2/2, 1 +i \nu_2, \bar{t}^2\Rb \right\}
\,\nonumber \\
&&\times \left\{ \,C^*_1(\nu_3) {}_2F_1\Lb 1/4 + i \nu_3/2, 3/4 + i \nu_3/2, 1 +i \nu_3, t^2 \Rb\,\,{}_2F_1\Lb 1/4 + i \nu_3/2, 3/4 + i \nu_3/2, 1 +i \nu_3, \bar{t}^2\Rb \right.\nonumber \\
&& \left.   - \,\,C^*_2(\nu_3)\Lb 4\,t\,\bar{t}\Rb^{2 i \nu_3}\, \,{}_2F_1\Lb 1/4 + i \nu_3/2, 3/4 + i \nu_3/2, 1 +i \nu_3, t^2 \Rb\,\,{}_2F_1\Lb 1/4 + i \nu_3/2, 3/4 + i \nu_3/2, 1 +i \nu_3, \bar{t}^2\Rb \right\}
\,\nonumber
\eea
with
\beq \label{B3}
C_1\Lb \nu \Rb\,\,=\,\,\frac{\pi^2}{ - i \nu}\,2^{2 i \nu}\,\frac{ \Gamma^2\Lb 1/2 - i \nu \Rb \,\Gamma\Lb i \nu \Rb}{\Gamma^2\Lb 1/2 + i \nu \Rb \,\Gamma\Lb - i \nu \Rb}\,;\,\,\,\,\,\,\,\,\,\,\,\,\,\,\,\,\,\,\,\,\,\,\,\,\,\,\,\,
C_2\Lb \nu \Rb\,\,=\,\,\frac{\pi^2}{ - i \nu}\,2^{-6 i \nu}\,\frac{ \Gamma^2\Lb 1 - i \nu \Rb \,\Gamma\Lb i \nu \Rb}{\Gamma^2\Lb 1 + i \nu \Rb \,\Gamma\Lb - i \nu \Rb}\,;
\eeq
For simplicity we  consider $\Gamma_{3P} \Lb q'_1,q'_2; \nu_1,\nu_2,\nu_3 \Rb $ in \eq{B2} for $q'_1=0$.

The integral of \eq{B3} has three region of potential divergency: $t \to 0$, $ t \to 1$ and $t \to \infty$.
  First,  let us study the behaviour of the integrant at $t \to \infty$. For doing this we  use  ( see  formula {\bf 9.132}(2)  of Ref. \cite{RY}) the following expression
\bea
{}_2F_1\Lb 1/4 + i \nu_2/2, 3/4 + i \nu_2/2, 1 +i \nu_2, t^2 \Rb &=&
\frac{\Gamma(1 + i \nu_2)\,\Gamma(1/2)}{\Gamma^2(3/4 + i \nu_2)} \Lb - \frac{1}{t^2}\Rb^{\frac{1}{4} + i \frac{\nu_2}{2}}\!\!\!\! \!\!\!{}_2F_1\Lb 1/4 + i \nu_2/2, 1/4 + i \nu_2/2, \frac{1}{t^2 }\Rb \label{B4}\\
  &+& \frac{\Gamma(1 + i \nu_2)\,\Gamma(-1/2)}{\Gamma^2(1/4 + i \nu_2/2)} \Lb - \frac{1}{t^2}\Rb^{\frac{3}{4 }+ i \frac{\nu_2}{2}} \!\!\!\!\!\!\!{}_2F_1\Lb 3/4 + i \nu_2/2, 3/4 + i \nu_2/2, \frac{1}{t^2 }\Rb \nonumber
\eea
From \eq{B4} we conclude that the integrant of \eq{B2} decreases at large $t$ as  $ t^{-2}$ for $\nu_2=\nu_3 =0$.
It means that integrals over  $t$ and $\bar{t}$ are convergent as far as large $t$ and $\bar{t}$ are concerned.
The next suspicious region is $t \,\to \,0$ ($\bar{t} \to 0$).
Factor $\Lb t\,\bar{t} \Rb^{-1/2 + i \nu_1 - i \nu_2 -  i \nu_3}$ in the integrant indicates a possible divergence in this region. Rewriting the integral in a general form
\bea \label{B5}
&&\Gamma_{3P} \Lb 0,q'_2; \nu_1,\nu_2,\nu_3 \Rb\,\,=\,\,\int^{\infty}_{0}\,\pi\,
d (t \bar{t})\,\Lb t\,\bar{t} \Rb^{-1/2 + i \nu_1 - i \nu_2 -  i \nu_3}\,\,\Phi(t,\bar{t})\,\,\\
&&=\,\,\int^1_0 d (t \bar{t})\,\Lb t\,\bar{t} \Rb^{-1/2 + i \nu_1 - i \nu_2 -  i \nu_3}\,\left\{ \Phi(0) + ( \Phi(t,\bar{t}) - \Phi(0)  ) \right\}
 \,\,+\,\,\int^{\infty}_1\,d (t \bar{t})\,\Lb t\,\bar{t} \Rb^{-1/2 + i \nu_1 - i \nu_2 -  i \nu_3}\,\,\Phi(t,\bar{t}) \nonumber
 \eea
we see that  if $\Phi(t, \bar{t}) - \Phi(0) \propto t\,\bar{t}$ at $ t \to 0$  we have a pole $1/(1/2 + i \nu_1 - i \nu_2 - i \nu_3)$.
However, since our function $\Phi(t, \bar{t})$ has  factors $t^{1/2 + i \nu_i}$ we need to be more  careful.  In calculating first enhanced diagram (see  section 3.4) $\nu_2$ and $\nu_3$ are small and  $\Phi( t, \bar{t} )$ has the following form
\beq \label{B6}
 \Phi( t, \bar{t} )\,\,=\,\,C_1(\nu_1) \frac{\pi^4}{- \nu_2\,\nu_3}\Lb 1 -  \Lb 4 t \bar{t} \Rb^{2 i \nu_2} \Rb \times \Lb 1 -  \Lb 4 t \bar{t} \Rb^{2 i \nu_3} \Rb\,\,\xrightarrow{\nu_2 \to 0,\, \nu_3 \to 0} \,\,C_1(\nu_1)\,4 \pi^4\,\,\ln^2\Lb 4 t \bar{t} \Rb
\eeq
Using \eq{B6} we obtain that
\beq \label{B7}
\Gamma_{3P} \Lb 0,q'_2; \nu_1,\nu_2,\nu_3 \Rb\,\,\xrightarrow{\nu_2 \to 0,\, \nu_3 \to 0} \,\,\frac{C_1(\nu_1)\,4 \pi^4\, \Lb \frac{{q'_2}^2}{4}\Rb^{1/2 - i \nu_1 + i \nu_2 + i \nu_3}}{(1/2 + i \nu_1 - i \nu_2 - i \nu_3)^3}\,\,+\,\,\mbox{function without singularities in $\nu_2$ and $\nu_3$}
\eeq

At $ t \bar{t} \, \to \,1$ we  could have a divergence which leads to a pole $1/(3/2 + i \nu_1)$.  However, since $C_1(\nu_1)$ has a zero of the second order at $i \nu_1 = -3/2$   $\Gamma_{3P} \Lb 0,q'_2; \nu_1,\nu_2,\nu_3 \Rb$ has no such singularity.
Therefore,  \eq{B6} gives the triple Pomeron at small $\nu_2$ and $\nu_3$. One can see that the  pole in \eq{B7}  replaces  the $\delta$-function which we obtained  integrating first of $q'_2$.

For arbitrary $\nu_2$ and $\nu_3$ the triple Pomeron vertex has a more complicated form, namely,
\bea
\Gamma_{3P} \Lb 0,q'_2; \nu_1,\nu_2,\nu_3 \Rb\,\,&=&\,\, C_1(\nu_1)\,4  \Lb \frac{{q'_2}^2}{4}\Rb^{1/2 - i \nu_1 + i \nu_2 + i \nu_3} \,\,\times  \,\,\left\{ \frac{C^*_1(\nu_2)\,C^*_1(\nu_3)}{ 1/2 + i \nu_1 - i \nu_2 - i \nu_3}\,\,-
\,\,\frac{C^*_1(\nu_2)\,C^*_2(\nu_3)}{ 1/2 + i \nu_1 - i \nu_2 + i \nu_3} \right. \label{B9} \\
& - &\left.\frac{ C^*_2(\nu_2)\,C^*_1(\nu_3)}{ 1/2 + i \nu_1 + i \nu_2 -  i \nu_3} \,\,+\,\,\frac{C^*_2(\nu_2)\,C^*_2(\nu_3)}{ 1/2 + i \nu_1 + i \nu_2 + i \nu_3} \right\} \,\,  + \,\,\mbox{regular function of $\nu_2$ and $\nu_3$} \nonumber
\eea

Integration over $q'_2$ leads $\delta( \nu - \nu')$ where $\nu $ ($\nu_2$ in \eq{loop-a-1}) corresponds to upper BFKL Pomeron in \fig{1endi}, and $\nu' $ ($\nu_4$ in \eq{loop-a-1}) corresponds to lower BFKL Pomeron in \fig{1endi}.

\end{document}